\documentclass[
 aip, jcp,
 amsmath,amssymb,amsfonts,
 preprint,
]{revtex4-2}

\usepackage{graphicx}
\usepackage{dcolumn}
\usepackage{bm}

\usepackage{hyperref}
\hypersetup{
    colorlinks=true,
    linkcolor=black,      
    urlcolor=blue,
    citecolor=black,
    }

\usepackage[version=4]{mhchem}
\usepackage{siunitx}
\usepackage[utf8]{inputenc}
\usepackage[T1]{fontenc}
\usepackage{mathptmx}
\usepackage{etoolbox}

\usepackage{textcomp, gensymb}

\makeatletter
\def\@email#1#2{
 \endgroup
 \patchcmd{\titleblock@produce}
  {\frontmatter@RRAPformat}
  {\frontmatter@RRAPformat{\produce@RRAP{*#1\href{mailto:#2}{#2}}}\frontmatter@RRAPformat}
  {}{}
}
\makeatother

\begin{document}


\title[Predicting Magnetic Janus Particle Assembly with Differential Evolution Algorithm]{Predicting Magnetic Janus Particle Assembly with Differential Evolution Algorithm}
\author{Eric A. McPherson}
 \affiliation{ 
 Department of Chemical Engineering, The City College of New York, CUNY, New York, NY 10031, USA
 }
 
\author{Kenneth Kroenlein}%
 \affiliation{ 
 Citrine Informatics, Redwood City, CA 94063, USA
 }

\author{Ilona Kretzschmar}
 \homepage{http://ikretzschmar.ccny.cuny.edu.}
 \affiliation{ 
  Department of Chemical Engineering, The City College of New York, CUNY, New York, NY 10031, USA
 }
 \email{kretzschmar@ccny.cuny.edu.}

\date{\today}

\begin{abstract}
Magnetic Janus particles allow access to complex, nonlinear assembled structures that may enable interesting new magnetorheological (MR) fluids with uniquely engineered field responses. 
However, the overwhelming size of the parameter space for Janus and patchy particles makes exploration of such systems by experimental trial and error or through detailed simulation impractical.
Here, a differential evolution (DE)-based simulation method is explored to predict the assembly of magnetic Janus particles as an alternative method for assembly prediction. 
Structure predictions from the DE simulation for laterally- and radially-shifted magnetic Janus particles are compared to four published experimental and simulation case studies.
The DE simulation captures the orientation and structure of magnetic Janus particles for a range of shifts and a variety of external field conditions using the point dipole approximation.
Structural predictions that rely on the reorganization of large clusters of particles were less well represented by the DE predictions. 
Despite this limitation, the DE simulation method can be used to predict key structural factors for magnetic Janus particle assemblies, as demonstrated by favorable comparison with three of the four model studies. 
\end{abstract}

\maketitle

\section{\label{sec:introduction} Introduction}
Magnetorheological (MR) fluids are colloidal suspensions that change their viscosity in the presence of a magnetic field.~\cite{ashtiani2015,eshgarf2022}
The viscosity change is caused by the assembly of magnetic microparticles due to magnetic dipole-dipole and dipole-field interactions.
The field-controlled viscosity of MR fluids has applications in mechanical dampers,~\cite{olabi2007} prosthetics,~\cite{jonsdottir2009} braking systems,~\cite{karakoc2008} bottom-up designable materials,~\cite{li2020a} and MR finishing for optical surfaces.~\cite{kumar2022} 
An MR fluid's response to a field must be characterized to take advantage of it in an engineering context; this necessitates understanding and predicting the assembly of the constituent particles. 

Janus particles are microparticles that have two different surface materials, giving the particles different functionalities including tunable responses to external fields.~\cite{chen2012, zhang2015, zhang2017, bishop2023}
Metallic and metal oxide coatings, including iron,~\cite{smoukov2009b} iron oxides,~\cite{ren2012} nickel,~\cite{sinn2011,demirors2018a} and composite caps made of cobalt/platinum layers,~\cite{suzuki1995, baraban2008} have been used to create particles that assemble in magnetic fields.
Their isotropic interactions have been modeled as a "shifted" magnetic dipole.~\cite{kantorovich2011}
The shifted dipole causes an increase in assembly complexity, expanding the range of potential structures and resultant macroscopic properties compared to magnetically isotropic particles.~\cite{tan2024}

Several research groups have explored the structures and potential applications of magnetic Janus particles with various cap designs.~\cite{zhang2015, zhang2017}
Smoukov \textit{et al.}~\cite{smoukov2009b} showed that magnetic Janus particles under the influence of magnetic and electric fields form chain structures that can then be demagnetized and disassembled.
Han \textit{et al.}~\cite{han2017b} made patchy magnetic microcubes that had programmed and manipulated folding shapes.
Our research group has fabricated ferro- and ferrimagnetic Janus particles and characterized their assembly structures and kinetics in an applied external field illustrating cap-material dependence of their assembled morphology.~\cite{ren2012, long2019}
While these investigations have led to a fundamental understanding of the assembly dynamics of magnetic Janus particles, experimental explorations of these materials are limited because of the time it takes to synthesize and test unique particles.
It is also nontrivial to characterize the as-manufactured magnetic properties of individual particles such as dipole shift, moment, and direction for magnetic caps.~\cite{philipp2021}

In parallel, simulations utilizing an array of particle types with well-defined physical properties have been performed \textit{in silico}.~\cite{eslami2019a, donaldson2021}
Kantorovich \textit{et al.}~\cite{kantorovich2011a} explored models of magnetic particles with radially-shifted dipoles using Monte Carlo simulations and found a wide range of structures for different shifts.
Vega-Bellido \textit{et al.}~\cite{vega-bellido2019} and Victoria-Camacho \textit{et al.}~\cite{a.victoria-camacho2020} used Brownian dynamics (BD) simulations to understand the dynamics of magnetic Janus particle assembly for laterally- and radially-shifted particles, respectively.
DeLaCruz-Araujo \textit{et al.}~\cite{a.delacruz-araujo2016a} used BD simulations to look at (nonmagnetic) Janus particles in shear flow, finding that they can assemble in a range of structures including micelles, vesicles, lamellae, and worm-like structures.
The simulations give insight to a larger structural complexity, show the impact of particle shape and Brownian motion, and validate the point-dipole approximation; however, it still takes significant time and computational resources to explore large particle systems with complex interactions.
In BD simulations this problem is further exacerbated as larger dipole shifts are explored because of the shorter time steps required.~\cite{vega-bellido2019}
Variations in size, shape, surface functionalization, and material of the particle core or cap also increase the complexity.
Further, external gradients (including chemical, electric, magnetic, and thermal gradients) impact the dynamics for directed self-assembly.~\cite{zhang2017}

These drawbacks motivate the search for alternative methods for predicting self-assembled structures.
Optimization strategies such as evolutionary algorithms can explore a vast configurational space quickly and predict particle structures without relying on detailed simulations.~\cite{deaven1995, daven1996} 
Evolutionary algorithms have been used to optimize many problems, including the assembly of colloidal crystals,~\cite{chremos2009} simulated patchy colloidal particles with Lenard-Jones interactions,~\cite{doppelbauer2010, bianchi2012} and cluster prediction for charged~\cite{cruz2016a} and Lenard-Jones colloids.~\cite{schonborn2009}
Evolutionary programming has been use by Brown \textit{et al.}~\cite{brown2010} to predict interatomic potentials, showing that algorithms can be parallelized. 
Inspired by biological evolution, these optimization methods are easy to interpret and parallelize.
Evolutionary algorithms are also widely applicable as long as the solution can be encoded as an array of values termed a "chromosome". 
Additionally, since they are heuristic-based optimization methods they have the potential to reach cluster predictions with fewer simulation steps and calculations.
Because of their ease of use and wide applicability, modifications have been made to create families of similarly related optimization methods. 
Differential evolution (DE) is a type of evolutionary algorithm~\cite{storn1997, ahmad2022} that has been shown to work better for continuous-value optimizations compared to genetic algorithms.~\cite{storn1997} 
Moloi and Ali~\cite{moloi2005} used DE as an algorithm to minimize potential energy functions for pair and many-body potentials and found that it worked favorably compared to other global minimization methods. 

This paper tests the ability of a DE-based simulation to predict magnetic Janus particle assembly. 
Key structure parameters are measured to compare the predictions from the DE simulation to a selected set of published experiments~\cite{ren2012, baraban2008} and simulations,~\cite{vega-bellido2019, okada2023} highlighting the advantages and limitations of the DE simulation.
This work first introduces the model systems that have been selected as test cases in Sec.~\ref{sec:models}. 
Details on the simulation and fitness function definition are provided in Sec.~\ref{sec:methods}.
In Sec.~\ref{sec:results}, the DE simulation results are introduced and compared to the model systems.
The main conclusions and future work for simulating magnetic Janus particles with DE are detailed in Sec.~\ref{sec:conclusions}.

\section{\label{sec:models}Model Systems}

\begin{figure}
    \includegraphics{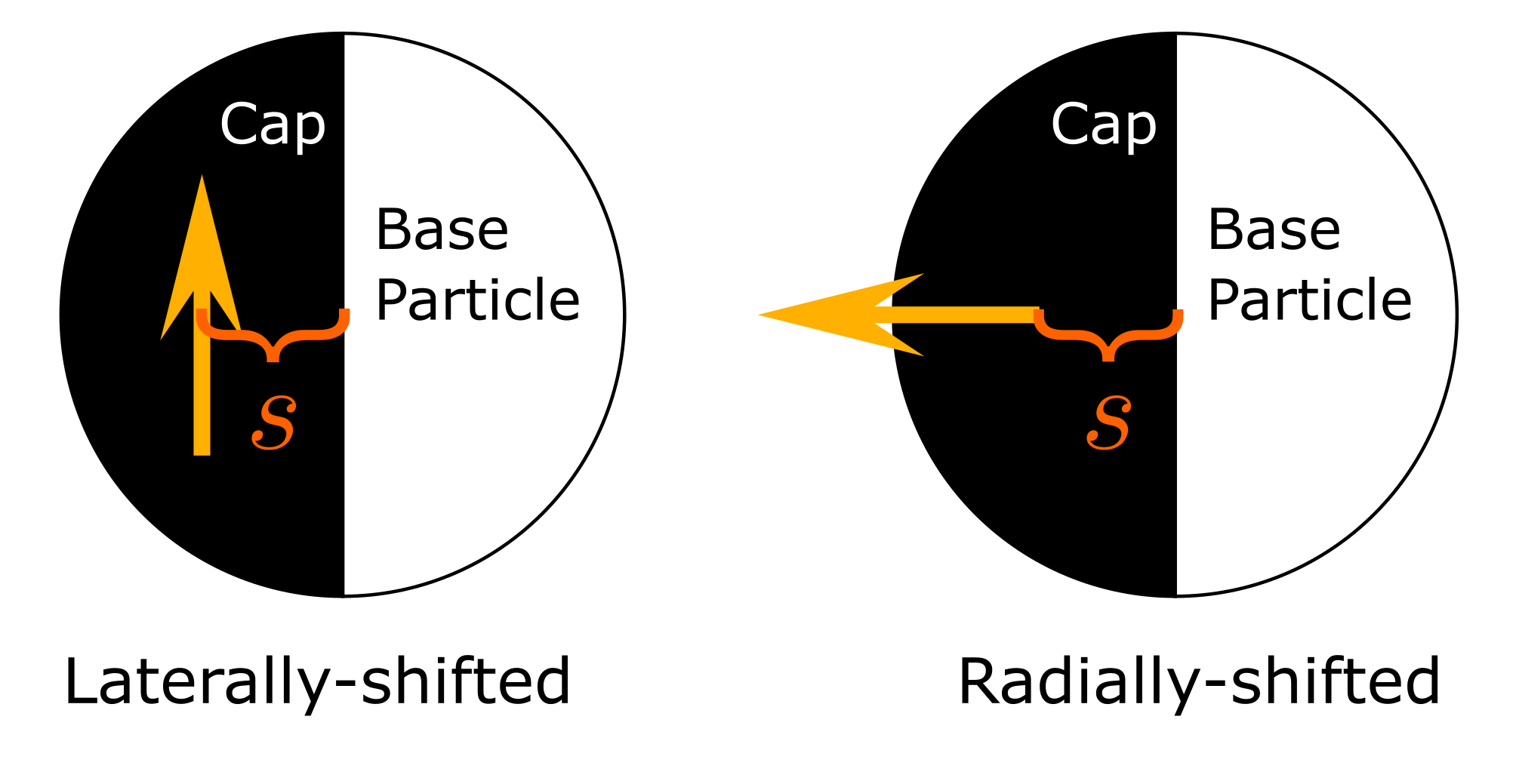}
	\caption{\label{fig:shift}Schematic comparing laterally- and radially-shifted dipole particles.
	The dipole shift $s$, normalized by the radius of the particle, is labeled for both types of particles.} 
\end{figure}

Magnetic Janus particles with a single patch fully coating one hemisphere are considered here.
Particles are assumed to have a single dipole that represents the magnetic moment of the cap material.
The dipole is modeled as being shifted away from the geometric particle center.
The distance of the dipole normalized by the particle radius is called the dipole shift, $s$, where the dipole is at the particle center when $s=0$ and the particle edge when $s=1$.
The dipole is shifted along the line between the particle's center and the cap's pole.
Figure~\ref{fig:shift} shows representations of the dipole shift for the two types of magnetic Janus particles studied here.
Iron- and iron oxide-capped particles are assumed to have their dipole oriented parallel to the cap equator, designated "laterally-shifted".~\cite{novak2015b,long2019}
Particles with alternating layers of cobalt and either palladium or platinum have their dipole pointing toward the pole of the cap, designated "radially-shifted".~\cite{a.victoria-camacho2020}

The DE simulation is compared against four published experimental and computational works exploring the self-assembly of laterally- and radially-shifted Janus particles, which are introduced in Sec.~\ref{sec:models}A-D. 
The model systems have been selected based on the complexity of observed structures, variety of study methods, and range of dipole and field conditions. 

\subsection{\label{sec:models_lateral_experimental}Magnetic Field Chain Formation}
Ren \textit{et al.}~\cite{ren2012} fabricated and then assembled laterally-shifted iron-oxide Janus particles suspended in water in a \SI{0.08}{T} magnetic field under quasi-two-dimensional conditions.
The oxidation state of the cap determines whether staggered, close-packed, or no chains formed.
The nearest-neighbor angle between pairs of particles in a chain and the external field, $\theta_{nn}$, was measured as $\theta_{nn} = 50\degree \pm 5 \degree$ and $\theta_{nn} = 59\degree \pm 5\degree$ for \ce{Fe_{1-x}O} and \ce{Fe3O4}, respectively, corresponding to staggered and double chains. 
The chain structure was attributed to the saturation magnetization for the different cap materials.
Cap thickness had no impact on chain structure, indicating that magnetization (and therefore dipole strength) alone do not fully explain the changes in structure.
Subsequent studies interpret the change in chain morphology using the concept of a shifted dipole, $s$.~\cite{novak2015b,long2019}

The chain-forming system is chosen because the magnetic field limits the structures that can form to two types of chains, which are characterized using $\theta_{nn}$.
A simulation method that replicates the results shows that it can capture the inter-particle interactions for a system that is energetically confined by an external field. 
Agreement would show that simplifying the magnetic cap as a shifted point-dipole is a valid model for describing magnetic Janus particle assembly using DE.

\subsection{\label{sec:models_radial_frustrated}Frustrated Clusters}
Baraban \textit{et al.}~\cite{baraban2008} made close-packed clusters of radially-shifted magnetic Janus particles.
The particles in the assembly had their magnetic dipoles fall in the same plane for certain "magic" numbers of particles, \textit{i.e.}, 3, 12, 27, and 48. 
The clusters were grown by adding single particles at a time using optical tweezers. 
The arrangement of the particle dipoles in the clusters matched the XY-spin on a triangular lattice (TL) model that confines magnetic dipoles to a TL in a single plane and assumes the dipoles are at the particles' center of mass.
Baraban \textit{et al.} also showed that some of the particles at the edge of the clusters diverged from ideal packing and arrangement because they had fewer nearest-neighbors to stabilize them. 
Further, the authors found that the TL could be tiled as a space filling pattern, where the internal particles matched the ideal pattern but the edge effects remained. 

The frustrated cluster system is chosen to test the DE simulation's ability to optimize the structure of closely packed particles with many neighbor interactions in the absence of an external magnetic field. 
The 12-particle system is a good test case because it contains both internal and external particles that allow the edge effects to be measured. 
A simulation method that replicates these results would be able to predict that internal particles line up with the idealized triangular lattice while external particles deviate from the lattice to better match their experimental micrographs.

\subsection{\label{sec:models_lateral_field}Changing Field Strength}

Quasi-2D Monte Carlo (MC) simulations reported by Okada and Satoh~\cite{okada2023} look at the effect of different magnetic field strengths on laterally-shifted Janus particle assemblies. 
They found that low magnetic fields resulted in small triplet- and quadruplet-clustering while increasing the magnetic field strength created longer and straighter chain structures. 
These results were captured by two order parameters, $S^{(e)}$ and $<n \cdot h>$, defined as 
\begin{equation}
    \label{eq:se}
	S^{(e)} = \frac{1}{N_{pair}} \langle \sum_{i=1}^{N} \sum_{j=1}^{N} P_2(\cos \psi_{ij}^{(e)}) \rangle
\end{equation}
and 
\begin{equation}
    \label{eq:nh}
	\langle \boldsymbol{n} \cdot \boldsymbol{h} \rangle = \frac{1}{N} \langle \sum_{i=1}^N (\boldsymbol{n_i} \cdot \boldsymbol{h}) \rangle
\end{equation}
where $P_2(\cos\psi^{(e)}_{ij})$ is the second Legendre polynomial, $\psi^{(e)}_{ij}$ is the angle between the poles of particles $i$ and $j$, $N$ is the total number of particles, $N_{pair}$ is the number of pairs of particles, and $j>i$.
The unit vector in the direction of the dipole moment for the $i$th particle is defined as $n_i$, while $h$ is the unit vector in the direction of the externally applied magnetic field.
The two order parameters characterized the particle-particle and particle-field orientations, respectively.
As the field strength increased, both order parameters increased, but  $\langle \boldsymbol{n} \cdot \boldsymbol{h} \rangle$ increased faster, while $S^{(e)}$ grew linearly.
The increases in order parameters correspond to the result that the particles aligned themselves with the field at low fields, while it took a stronger field to have particles be more aligned with themselves and form straighter chains.

The field strength was nondimentionalized as $\xi = \mu_0 m H / (kT)$, where $\mu_0$ is the permeability of free space, $m$ is the magnitude of the magnetic dipole, $H$ is the magnitude of the magnetic field strength, $k$ is Boltzmann constant, and $T$ is the absolute temperature of the system.
The dipole strength was nondimentionalized using the dipolar coupling constant $\lambda=\mu_0 m^2 / 4 \pi d^3 k T$, where $d$ is the particle diameter.

The changing field strength system is chosen to test the predictions of minor changes to cluster structure caused by the interplay of particle-particle and particle-field interactions by the DE simulation.
Since $\xi$ and $\lambda$ are both nondimentionalized by $kT$, their ratio $\xi/\lambda = \mu_0 m/4 \pi d^3 B$, where $B$ is the magnetic flux density from using the relationship $B=\mu_0 H$, is independent of temperature.
The DE simulation does not include a temperature, so $\xi / \lambda$ can be used to compare the DE results to the published work. 
A simulation method that captures the subtle interplay between dipole-dipole and dipole-field interactions will predict similar trends in the order parameters.
Comparisons of the order parameters test ability for the DE simulation to predict structures in the absence of Brownian motion. 

\subsection{\label{sec:models_lateral_ubaldo}Non-directed Assembly}

BD simulations by Vega-Bellido \textit{et al.}~\cite{vega-bellido2019} explored the assembly of laterally-shifted magnetic Janus particle as a function of time and dipole shift without an external magnetic field.
Particles with lower shifts were found to have a time dependency on cluster growth rates because they formed chain and loop structures that continued to grow.
The nucleation factor $n_c$ characterizes cluster size and is defined as: 
\begin{equation}
	n_c = \frac{N_c - N_s}{N_p}
\end{equation}
where $N_c$ is the number of clusters (including singlets), $N_s$ is the number of singlets, and $N_p$ is the total number of particles. 
The nucleation and growth of the clusters can be followed when $n_c$ is measured throughout the simulation.
The simulations showed that the formation of clusters broadly splits into two phases: first, in the nucleation phase, $n_c$ rapidly increased as the singlet particles form small clusters (doublets and triplets). 
Second, in the growth phase, the nucleation factor decreased as the remaining singlets and small clusters group together into larger clusters.
While the simulations showed that $n_c$ increased at a similar rate for all dipole shifts simulated, the decrease in nucleation factor in the growth phase was much slower for higher shifts as they tended to form small, dense clusters.

The non-directed self-assembly system is chosen to test if the DE simulation converges on a path similar to the BD simulation and reproduces the distinct nucleation and growth phases observed, indicating DE's ability to model cluster rearrangement in less restricted systems.
Similar to the changing field strength case study in Sec.~\ref{sec:models}C, comparisons of the nucleation factor also test the ability for the DE simulation to predict structures in the absence of Brownian motion.

\section{\label{sec:methods}Methods}
A DE simulation is written to predict the assembly of laterally- and radially-shifted magnetic Janus particles.
The simulation is based on a published Python example~\cite{cristina2021} that has been significantly modified and vectorized to take advantage of the Numpy Python package optimizations.~\cite{harris2020}
Particle arrangements are encoded as a chromosome, which is used to generate new trial solutions.
DE reproduction and random gene mutation are used to generate the trial solutions.
A fitness function based on the system's magnetic potential energy is minimized using DE.
The starting arrangement is randomly generated, and a preconditioning step is first run to remove any overlapping particles before starting the DE simulation.

\subsection{\label{sec:methods_de}Differential Evolution}

The algorithm used in this work follows the same structure as Storn and Price~\cite{storn1997} with the additional use of a random mutation factor, typically used in genetic algorithms.~\cite{katoch2021b} 
Trial solutions to the problem, known as individuals, are encoded into vectors called chromosomes.
An initial population of individuals is randomly generated at the start of the algorithm, and genetic operations are used to generate trial solutions from the previous generation to evolve toward a more optimized population over a predefined number of generations. 
The number of individuals in a population is called the population size.
The $i$th variant solution for the next generation, $p^v_i$, is generated with the relationship: 
\begin{equation}
	p^{v}_{i} = p_j + c_F\cdot (p_k - p_l)
\end{equation}
where $p_j$ is the $j$th chromosome in the population, $i, j, k,$ and $l$ are all random and different indices, and $c_F$ is the mutation factor. 
Finally, a chance of random mutation occurs using a different uniformly distributed random number $k_i $: if $k_i < c_{mut}$, the random mutation parameter, a single gene in the chromosome is randomly selected and changed to a randomly generated value. 
Note that crossover is omitted to simplify the DE algorithm ensuring that it works for the wide variety of case studies explored here.~\cite{zhang2023}
To increase the exploration space, the variant solution is accepted as a member of the next generation by using a uniformly distributed random number $r_i$ in the relationship:
\begin{equation}
	\label{eq:crossover}
	p^t_i = 
            \begin{cases}
				p^v_i, &\quad\text{if } r_i < c_R \quad\text{and } f(p^v_i) < f(p_i) \\
                p_i, &\quad\text{otherwise}
            \end{cases} 
\end{equation}
where $p_i$ is the $i$th individual in the current generation, $c_R$ is the acceptance parameter, $p^t_i$ is the $i$th member of the next generation, and $f(p)$ is the fitness of chromosome $p$. 
The chromosomes have their values clipped to between [-1, 1] to ensure they are valid trial solutions.	

\subsubsection{\label{sec:methods_chromosome}Chromosome Encoding}

\begin{figure}
    \includegraphics{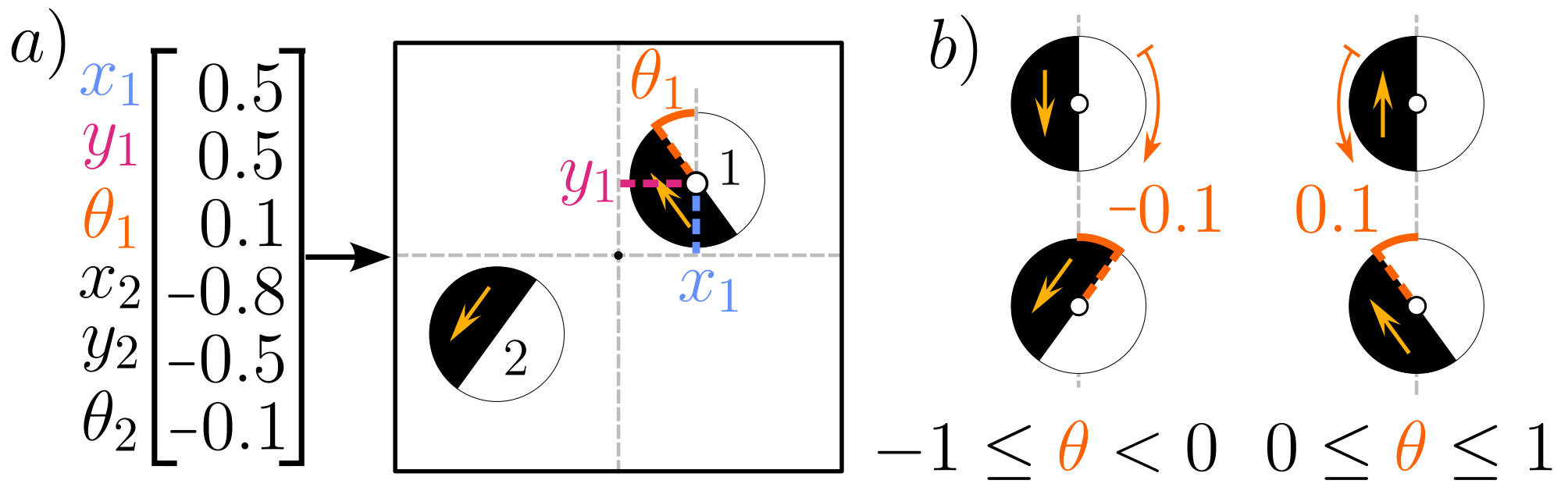}
	\caption{\label{fig:encoding}Representation of a) chromosome encoding an arrangement of particles, and b)$
\theta_{i}$ gene encoding both in-plane rotation and mirroring/flipping for laterally-shifted particles. Each gene represents either a position in 2D ($x_{i}, y_{i}$) or a rotation ($\theta_{i}$). Radially shifted particles have axial symmetry and do not need this adjustment. All genes are normalized to fall between -1 and 1, inclusive. 
}
\end{figure}

The chromosomes in the DE simulation encode arrangements of magnetic Janus particles in a two-dimensional box, as represented in Figure~\ref{fig:encoding}.
Three genes from each individual's chromosome are needed to represent a particle, as shown in Figure~\ref{fig:encoding}a, so each chromosome is $3N$ in length, where $N$ is the number of Janus particles. 
The first two genes represent the $x$ and $y$ positions of the particle, normalized so that -1 and 1 correspond to the edges of the box. 
The third gene represents the angle of rotation from the vertical axis, normalized so that zero represents the dipole pointing up, and -1 and 1 corresponding to a full counterclock- and clock-wise rotation, respectively. 
To account for the asymmetry of the laterally-shifted Janus particles, the out-of-plane rotation also needs to be encoded.
The rotation gene accounts for this aspect by mirroring the particle and rotation direction depending on the sign of the value.
A positive rotation gene means the dipole is facing up when the cap is on the left side of the particle and rotates counterclockwise away from zero, and a negative rotation gene means that the dipole is facing down when the cap is on the left and rotates clockwise away from zero, as illustrated in Figure~\ref{fig:encoding}b.

\subsubsection{\label{sec:methods_fitness}Fitness Function}
A fitness function determines which of the individuals are more fit using the chromosome as an input and is the objective function DE attempts to minimize.
For the simulations, the fitness score is essentially the potential energy of the system due to the magnetic interactions and consists of three parts: the dipole-dipole interactions ($U_{int, ij}$), dipole-field interactions ($U_{field, i}$), and the hard-sphere potential ($U_{overlap, ij}$), defined in equations~\ref{eq:dipole_dipole_potential}-~\ref{eq:intersections}.
\begin{equation}
	\label{eq:dipole_dipole_potential}
	U_{int,ij} = -\frac{\mu_0}{4\pi |\boldsymbol{r_{ij}}|^3} [3(\boldsymbol{m_i} \cdot \boldsymbol{\hat{r}_{ij}})(\boldsymbol{m_j} \cdot \boldsymbol{\hat{r}_{ij}}) - \boldsymbol{m_i}\cdot \boldsymbol{m_j}]
\end{equation}

\begin{equation}
	\label{eq:dipole_field_potential}
	U_{field,i} = -\boldsymbol{m_i} \cdot \boldsymbol{B}
\end{equation}

\begin{equation}
	\label{eq:intersections}
	U_{overlap,ij} = 
            \begin{cases}
                \infty, &\quad\text{if } d_{ij} < 2 \cdot r_{particle} \\
                0,       &\quad\text{otherwise}
            \end{cases} 
\end{equation}
Here, $\boldsymbol{r_{ij}}$ is the distance vector between the dipoles of the $i$th and $j$th particles, $\boldsymbol{\hat{r}_{ij}}$ is the distance unit vector, and $d_{ij}$ is the distance between the $i$th and $j$th particle centers. 

\subsection{\label{sec:methods_preconditioning}Preconditioning}

A preconditioning step is run before the main DE simulation because some particles may overlap when the initial population is randomly generated.
Running the preconditioning step significantly reduces convergence time.
The preconditioning step moves particles away from each other proportional to the amount of overlap along the axis formed between the centers of the two particles.
For the concentrations used in this paper, most preconditioning steps need less than 10 iterations. 

\subsection{\label{sec:methods_parameters}DE Parameters}
All simulations reported are done with 100 particles with the exception of the frustrated clusters, which only have 12 particles. 
Testing of up to 200 particle systems show no significant change (see supplementary material, Fig. S1).  
Figure S1 also shows that for more particles, more generations are needed to reach the same optimization point. 
Each data point is derived from 30 repeat simulations, and error bars show one standard deviation.
The physical system parameters are shown in supplementary material, Table S1.

Simulations are done changing the DE parameters for each case study to find the combination that gave the best fitness values.
It is found that there are relatively wide combinations of parameters that give similar results. 
Here, all simulations are run for $10^5$ generations and have a population size of 30. 
DE simulations were run with $c_F$, $c_{mut}$, and $c_R$ as 0.5, 0.78, and 0.91, respectively, unless otherwise noted.

\section{\label{sec:results}Results and Discussion}

\subsection{\label{sec:results_magnetic_chain_formation}
Results for Magnetic Field Chain Formation}

\begin{figure}
    \includegraphics[]{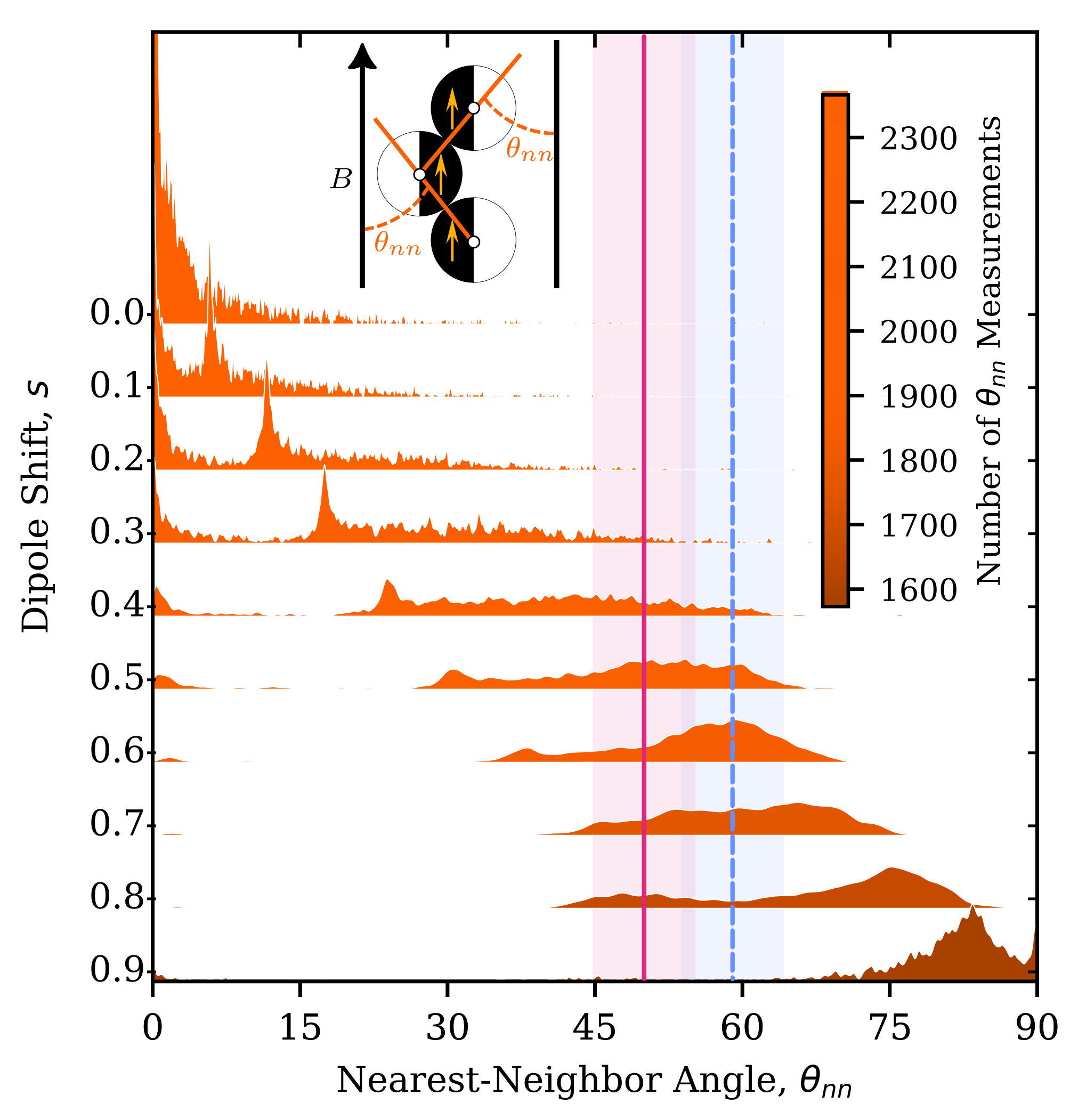}
	\caption{\label{fig:experimental}
    Kernel density estimation (see text) of distributions of nearest-neighbor angles $\theta_{nn}$ as a function of shift $s$ for chains with at least two particles from DE simulations.  
    Solid red and dashed blue lines represent the experimentally measured $\theta_{nn}$ and their uncertainty for \ce{Fe_{1-x}O} ($\theta_{nn} = 50\degree \pm 5\degree$) and \ce{Fe3O4} ($\theta_{nn} = 59\degree \pm 5\degree$), respectively, as reported by Ren \textit{et al.}~\cite{ren2012}
    The inset defines $\theta_{nn}$. 
    Curve color and color bar indicate the total number of $\theta_{nn}$ measurements from all simulations for each $s$.
    }
\end{figure}

The nearest neighbor angle ($\theta_{nn}$) is evaluated for the structures predicted by the DE simulations with $s=0.0$ to $0.9$ in increments of 0.1.
Figure~\ref{fig:experimental} shows the kernel density estimation (KDE) of $\theta_{nn}$ distributions at each $s$ for chains with at least two particles.
The normalized KDE uses a Gaussian kernel and the bandwidth for each shift is determined using the Improved Sheather-Jones algorithm~\cite{botev2010} in the KDEpy package.~\cite{2025} 
A narrow peak is observed at $\theta_{nn}\approx 0\degree$ for $s=0.0$ indicating a predominance of linear chains that persists at higher $s$, likely as a result of the DE simulation not breaking or reorienting chains after they assemble.
A second, sharp peak appears at $\theta_{nn}\approx 6\degree$ for $s=0.1$ that moves to larger angles from $s=0.2$ ($\theta_{nn}\approx 12\degree$) and $s=0.3$ ($\theta_{nn}\approx 17.5\degree$).
The peak decreases with its center at ($\theta_{nn}\approx 24\degree$) for $s=0.4$ and a second, much broader $\theta_{nn}$ distribution is observed ranging up to $\theta_{nn}\approx 65\degree$. 
This pattern repeats for $s=0.5$ with the peak (now at $\theta_{nn}\approx 31\degree$) further decreasing and the maximum of the distribution moving to $\theta_{nn}\approx56 \degree$.
At $s=0.6$, the lower peak has almost disappeared, whereas a pronounced narrowing of the broad distribution with a center at $\theta_{nn}\approx 60\degree$ occurs.
At $s=0.8$, a bifurcation in the $\theta_{nn}$ distribution appears, with peaks at $\theta_{nn}\approx49\degree$ and $\theta_{nn}\approx75\degree$.
The bifurcation corresponds to $\theta_{nn}$ approaching the characteristic nearest-neighbor orientations for colloidal square packing of $45\degree$ and $90\degree$.
The lower peak is no longer present for $s=0.9$, which indicates that most clusters are side-by-side doublets with a negligible amount of chains forming.
This finding is also indicated in the lower number of $\theta_{nn}$ observations made for the $s=0.9$ simulations (darker colored KDE) compared to lower $s$ (lighter colored KDE); i.e., fewer chains provide fewer $\theta_{nn}$ angle measurements.
Note that the $\lambda$ used is uncorrected for magnetically shifted particles;~\cite{klinkigt2013, weeber2013} therefore, the particle radius affects the assembly, as seen in supplementary material, Figure S2.

The red and blue lines in Figure~\ref{fig:experimental} represent the experimental $\theta_{nn}$ values measured for staggered and double chains, respectively, by Ren \textit{et al.}~\cite{ren2012}
The peak of the $\theta_{nn}$ distribution for $s=0.6$ coincides with the experimental double chain $\theta_{nn}$ (blue, dashed line) suggesting that Janus particles with \ce{Fe3O4} caps may be best described with this shift.
To find the range of potential $s$ that give $\theta_{nn}$ values correlating with the \ce{Fe3O4} system, additional simulations are run with an $s$ step-size of 0.02 for $s = 0.4 - 0.6$ (see supplementary material, Fig. S3).
The detailed simulations show that systems with an $s$ as low as $s\approx0.56$ can match the experimental observations for \ce{Fe3O4}.
It also allows correlation for the relationship between the DE predictions and experimental measurements for particles with \ce{Fe_{1-x}O} caps.
The distribution shows a maximum close to the experimentally-observed $\theta_{nn}\approx 50\degree$ at $s\approx0.48$, indicating that the experimentally observed staggered chains found for \ce{Fe_{1-x}O} caps may have this shift.

The $\theta_{nn}$ predictions in Figure~\ref{fig:experimental} are compared to $\theta_{nn,min}$ at which Eqn.~\ref{eq:dipole_dipole_potential} returns a minimum energy value to shed light on the $\theta_{nm}$ distributions and positions observed (see supplementary material, Fig. S4). 
Note, in an external field of $B=0.08 T$ used by Ren \textit{et al.},~\cite{ren2012} the magnetic dipoles are expected to perfectly align with the external field, simplifying Eqn.~\ref{eq:dipole_dipole_potential} by setting $\boldsymbol{m_i} \cdot \boldsymbol{m_j} = 1$.
The DE simulation predictions for $s=0.0$ and $s=0.6 - 0.9$ agree with the theoretical predictions, whereas $\theta_{nn}$ for $s=0.1 - 0.5$ diverges from $\theta_{nn,min}$ but instead match the angle found when both magnetic dipoles are collinear. 
It is unclear why the DE simulation would prefer a vertical alignment of the magnetic dipoles because the simplified Eq.~\ref{eq:dipole_dipole_potential} does not show a minimum at this angle. 
The finding may hint at the impact of out-of-plane rotation restriction in the DE simulations. 
The unavailability of experimental and simulation data for magnetic Janus particles with $s=0.1-0.4$ in strong external magnetic fields prevents further resolution of the finding at this point. 

\subsection{\label{sec:results_frustrated}Results for Frustrated Clusters}
\begin{figure}
\includegraphics{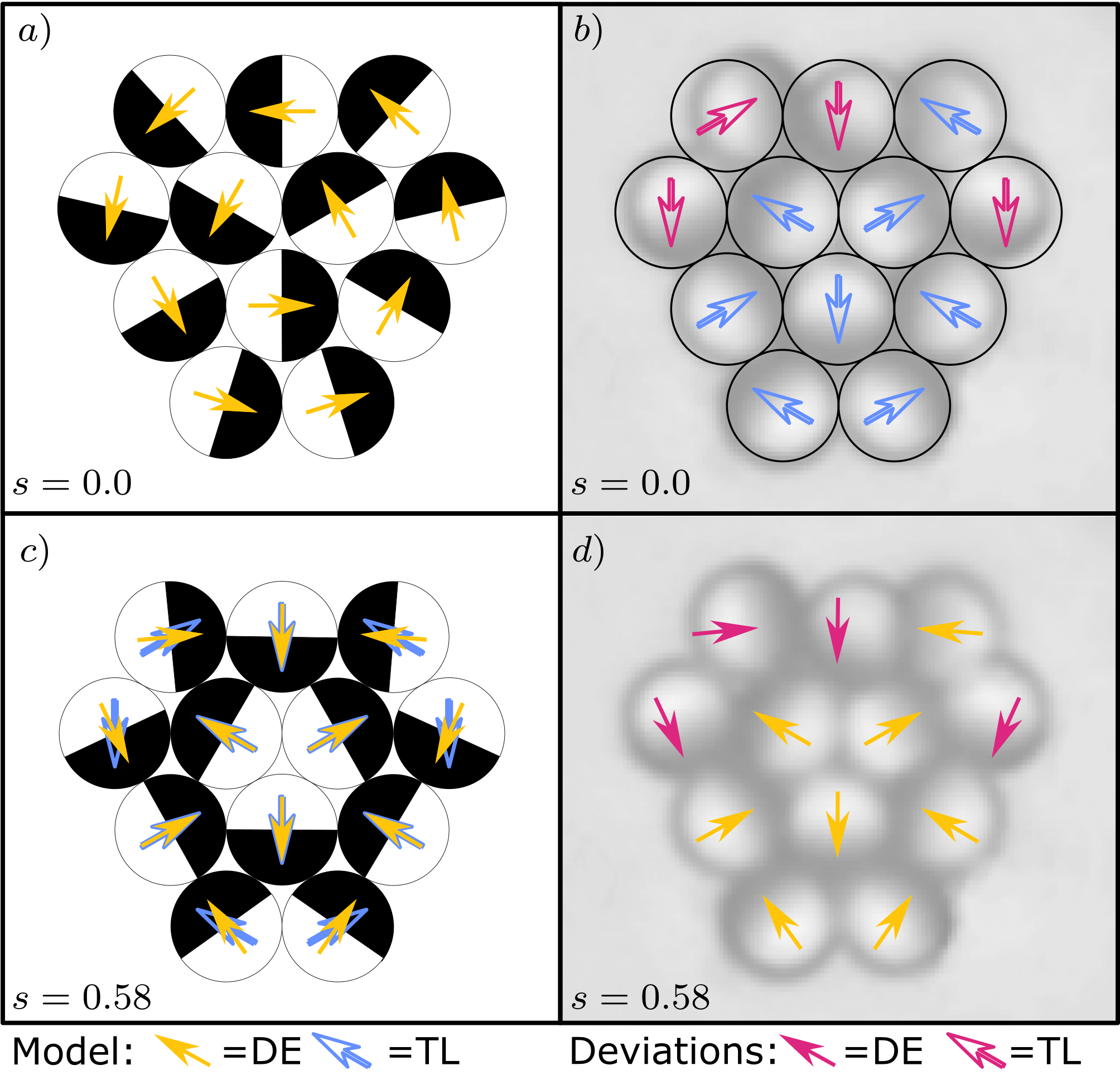}
\caption{\label{fig:frustrated} DE predictions (orange solid arrows) compared to experimental micrograph and triangular lattice (TL) model (blue hollow arrows) proposed by Baraban \textit{et al.} for radially-shifted Janus particle orientations in 12-particle magic clusters. 
a) DE simulation for $s=0$ with $f = 0.01 \pm 0.04$.
b) Overlay of TL model and experimental micrograph from Baraban \textit{et al.} showing the expected deviations of edge particles (red hollow arrows).
c) DE simulation for $s=0.58$ with $f = 0.94 \pm 0.05$.
d) Overlay of DE simulation orientations and experimental micrograph from Baraban \textit{et al.}, showing that edge deviations are predicted by the DE model (red solid arrows).
DE simulated structures and dipole orientations shown are representative results close to the median $f$-score (see supplementary material, Figs. S5 and S6). 
Red hollow and solid arrows show deviations in alignment for the experimental clusters and the TL and DE models, respectively (see text).
Micrograph used in parts b and d adapted with permission from~\textcite{baraban2008}
Copyrighted by the American Physical Society
}
\end{figure}

The DE simulations generally retain the 12-particle cluster structure for radially-shifted, magnetic Janus particles when the simulations start with the particles in the magic cluster positions reported by Baraban \textit{et al.} with initial orientations that match the triangular lattice (TL) model and the particles are allowed to relax and reorient during the simulation. 
In contrast, when the DE simulations start with particles in random orientations on the TL and are allowed to relax, the 12-particle cluster structure is rarely maintained. 

Figure~\ref{fig:frustrated} summarizes the DE predictions and compares them to the TL model and experimental micrographs. 
The DE simulation run with $s=0$, assumed by the TL model, results in head-to-tail aggregation, as shown in Figure ~\ref{fig:frustrated}a, in contrast to both the TL model and experimentally observed cap orientations shown Figure~\ref{fig:frustrated}b.
A similarity score, $f$, is introduced to quantitatively compare the DE simulation dipole orientation to the TL model, defined as:
\begin{equation}
	f = \frac{1}{12} \sum_{i=0}^{12} \cos{\theta_{diff,i}}
\end{equation}
where $\theta_{diff, i}$ is the angle between the DE simulation and the respective TL orientation for particle $i$.
Here, $f=1$ means the DE prediction exactly aligns with the TL model, while $f=-1$ when the DE prediction is antiparallel.
For $s=0.0$, $f = 0.01 \pm 0.04$, indicating poor agreement between the DE simulation and TL model, as seen in Figure~\ref{fig:frustrated}a and b.

Next, a sweep is performed over shifts from $s=0$ to 0.98 with a step size of 0.02.
The cluster obtained for a shift of $s=0.58$ most closely aligns with the TL model with an $f$-score of $f = 0.94 \pm 0.05$.
This result agrees with MC simulations by Kantorovich \textit{et al.},~\cite{kantorovich2011a} who found that an $s=0.6$ aligned closest to the TL model.
Figure~\ref{fig:frustrated}c shows a representative DE-predicted structure for $s=0.58$ overlaid with the TL model.
The interior particles align with the TL model and the second shell edge particles show the expected misalignment mentioned in Sec.~\ref{sec:models_radial_frustrated}. 

Figure~\ref{fig:frustrated}d shows the overlay of the DE predicted structure in Figure~\ref{fig:frustrated}c with $s = 0.58$ on top of the experimental micrograph from Baraban \textit{et al.}, also showing good agreement.
The red arrows in Figure~\ref{fig:frustrated}b and d indicate deviations between the models and experimental positions of the particle caps that may arise from non-uniform or out-of-plane particles. The DE simulation dipole orientations (red solid arrows) align closely with the experimentally observed cap orientations, indicating that the DE simulation method predicts the edge effects observed by Baraban \textit{et al.}~\cite{baraban2008}

\subsection{\label{sec:results_lateral_field}Results for Changing Field Strength}
\begin{figure}
\includegraphics{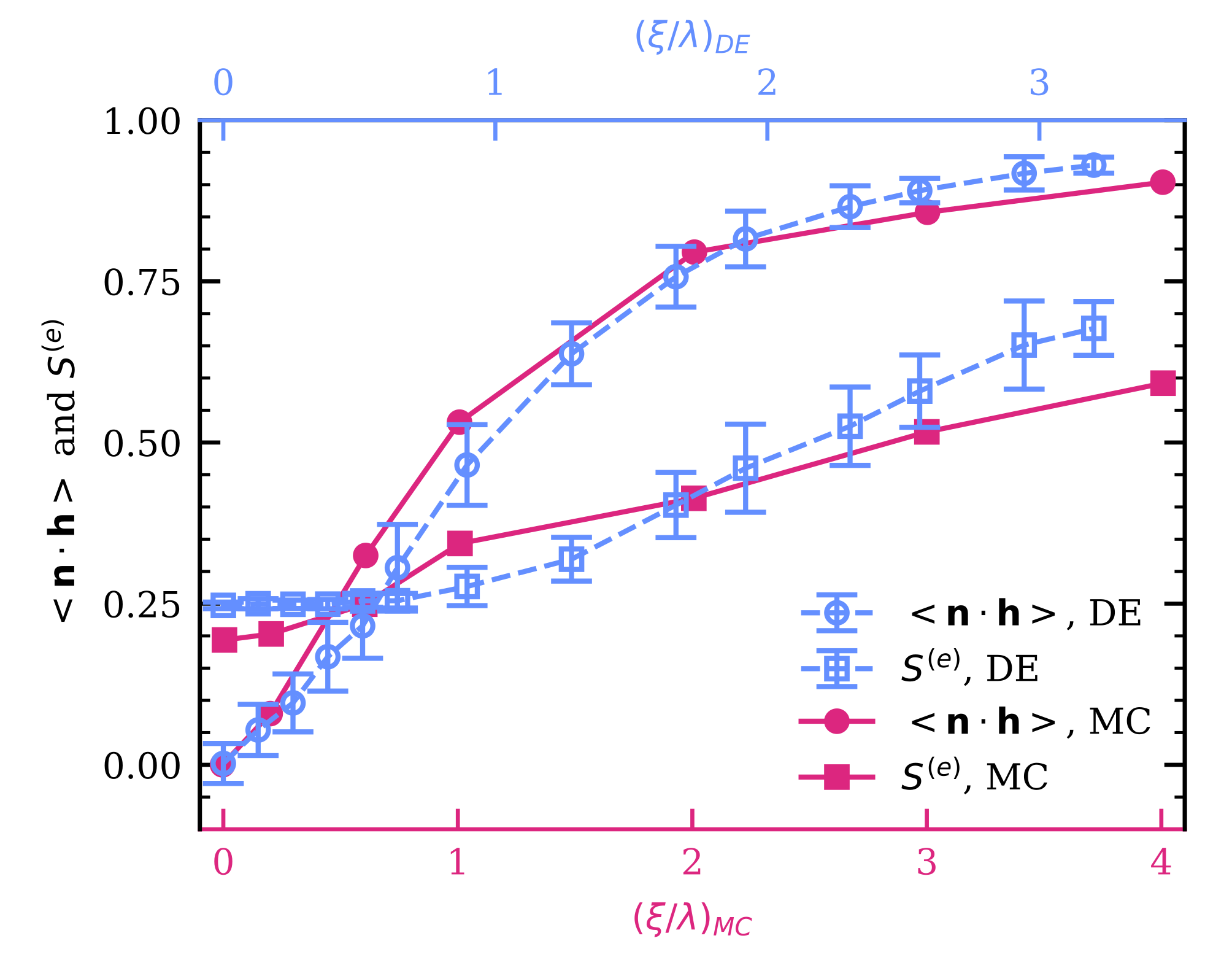}
\caption{\label{fig:field}Change in order parameters $S^{(e)}$ (squares) and $\langle \boldsymbol{n} \cdot \boldsymbol{h} \rangle$ (circles) as a function of external magnetic field at constant $ \lambda = 5$.
Parameters are plotted against $(\xi / \lambda)_{MC}$ for the MC simulation data from Okada and Satoh~\cite{okada2023} (red, solid line and symbols) and $(\xi / \lambda)_{DE}$ for the DE simulation (blue, dashed line and open symbols).
The scaling between the top, blue and bottom, red axes is 1.16.
Data from Okada and Satoh~\cite{okada2023} read using an online tool.~\cite{2024}
Reprinted by permission from Copyright Clearance Center.
}
\end{figure}

The DE simulations are run with magnetic field strengths from 0 to \SI{2.0e-4}{T} and the order parameters $\langle \boldsymbol{n} \cdot \boldsymbol{h} \rangle $ and $S^{(e)}$, defined in Sec~\ref{sec:models_lateral_field}, are determined. 
Owing to the higher volume fraction (see supplementary material, Table S1), a smaller mutation factor $c_F$ of \SI{1e-2} is used to generate non-overlapping children and increase the algorithm's exploration.

Figure~\ref{fig:field} shows the change in $\langle \boldsymbol{n} \cdot \boldsymbol{h} \rangle$ and $S^{(e)}$ for the final structures of the MC simulations by Okada and Satoh, $(\xi / \lambda)_{MC}$ (red, bottom axis), and the DE simulations, $(\xi / \lambda)_{DE}$ (blue, top axis).
The scaling between the two axis is \SI{1.16e0} found using a least-squares optimization. 
Results for the $c_F = 0.5$ case (supplemental material, Fig. S7) have larger standard deviations and require a larger scaling factor of \SI{1.45}.
The DE simulations show a similar trend in order parameters as the MC simulations but a weaker magnetic field is needed to predict the same order parameters in the DE simulation. 
In the absence of a magnetic field, changing the dipole strength $|\boldsymbol{m}|$ for the particles has no impact on the structures that form since the only energies present in the DE simulation without overlapping particles are the dipole-dipole and dipole-field interactions (Eqs.~\ref{eq:dipole_dipole_potential}  and~\ref{eq:dipole_field_potential}).
The structures that form are a result of the interplay between these two interactions.
Without the dipole-field interactions, changing $|\boldsymbol{m}|$ only serves to scale the systems energy without changing the results.

Both simulation methods use the same potential energy functions and neglect Brownian motion.
Therefore, the proportional difference could instead be due to a difference in the simulation techniques themselves. 
A missing phase of cluster growth and rearrangement could lead to smaller clusters that would be more strongly influenced by a magnetic field.
Additionally, shorter chains would not be able to bend like the longer chains in the MC simulation. 
The missing cluster growth and rearrangement will be further explored in Sec.~\ref{sec:results_nucleation_factor}.

\subsection{\label{sec:results_nucleation_factor}Results for Non-directed Assembly}

\begin{figure}
	\includegraphics[]{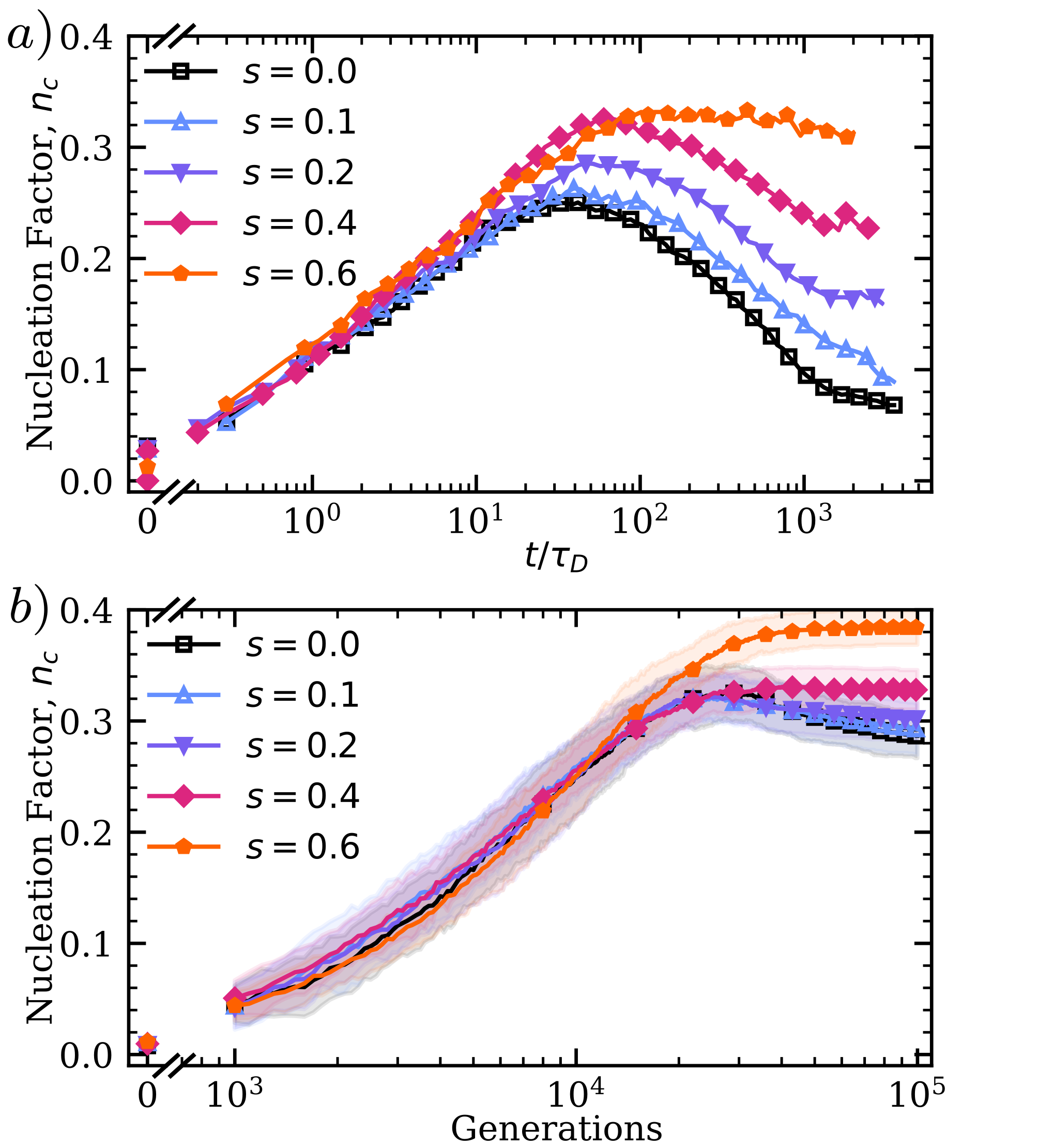}
	\caption{\label{fig:nucleation_factor} Change in nucleation factor $n_c$ over simulation time for 
	a) BD Simulations reproduced from Vega-Bellido \textit{et al.},~\cite{vega-bellido2019}
b) DE simulation (time represented by DE generations).
Colors and symbols represent the same dipole shifts in both simulations.
The shaded area represents one standard deviation from the average of 30 runs each. 
Standard deviations for a) are minimal and not shown.
The $n_c$ for the DE simulations are calculated every 100 generations.
For both figures, symbols are placed every few data points for readability.
An online graph-reading tool~\cite{2024} was used to read the data from Vega-Bellido \textit{et al.}~\cite{vega-bellido2019}
Reprinted by permission from Copyright Clearance Center.
}

\end{figure}
The DE simulations are run with laterally-shifted particles without an external field to track cluster nucleation and growth over DE simulation generations.
Figure~\ref{fig:nucleation_factor} compares the nucleation factor $n_c$ for the DE simulations (Fig.~\ref{fig:nucleation_factor}a) and the BD simulation by Vega-Bellido \textit{et al.}~\cite{vega-bellido2019} (Fig.~\ref{fig:nucleation_factor}b).
Note that the "time" axis is represented by generations for the DE simulations and time $t$ is nondimensionalized by the characteristic diffusion time of the particles, $\tau_D$, for the BD simulations. 
While the generations used in the DE simulations are not the same as $\frac{t}{\tau_D}$, both can be used as a way to quantify simulation progress. 
Therefore, the trends are more important than the actual generation and $\frac{t}{\tau_D}$ values.
The $n_c$ for the DE simulation reveals a similar nucleation phase, indicating both simulation methods start with nucleating singlets into small clusters. 

However, there is a distinct lack of a clear growth phase in later generations of the DE simulation, which only show a minor growth phase for lower shifts, and none for high shifts. 
The missing growth phase could be due to the algorithm being unable to move the clusters once they have formed; in other words, it is too energetically unfavorable to break up the clusters for the chance to reassemble them in a larger and ultimately more energetically favorable structure. 
Instead, the simulations enter a "frozen" phase where the population becomes homogeneous and random mutation alone is unlikely to move the system toward a more optimized outcome.
The higher peak in $n_c$ for the DE simulation is also explained by the difficulty to move large clusters, as there is likely a small amount of growth occurring in the nucleation stage of the BD simulation that is missing in the DE simulation.
While the growth phase is not observed, the DE simulation still captures some key aspects of the system's $n_c$. 
For example, the rank order of the average $n_c$ for the different shifts is preserved, with the no-shift particles ($s = 0$) having the lowest final $n_c$ and the high-shift particles ($s = 0.6$) having the highest.
This finding indicates that the underlying physics of the assembly is being captured despite the missing growth phase.
The lack of a growth phase can also help explain the scaling factor needed in Sec.~\ref{sec:results_lateral_field}.

\section{\label{sec:conclusions}Conclusions}

\textit{A priori} prediction of structures formed by anisotropic magnetic colloids is needed to develop new MR fluids with uniquely engineered flow properties. 
However, progress has been hampered by the time needed to explore the available parameter space.
Evolutionary algorithms are capable of searching a vast parameter space but have been explored only minimally for colloids.
Here, a DE-based simulation method is introduced and tested as a tool to predict the assembly of laterally- and radially-shifted magnetic Janus particles. 
The DE simulation viability is quantitatively explored using several model systems. 
The DE simulation is shown to accurately predict particle orientations for both particle-particle and particle-field magnetic interactions for systems with a strong and weak externally aligning field (Sec.~\ref{sec:results_magnetic_chain_formation} and Sec.~\ref{sec:results_lateral_field}) and system without fields (Sec.~\ref{sec:results_frustrated}).
The accurate orientation predictions are also made for a range of shifts (Sec.~\ref{sec:results_magnetic_chain_formation}), and are used to interpret experimental results.

The DE simulation has been used to predict the shift for staggered \ce{Fe_{1-x}O}- and double \ce{Fe3O4}-capped laterally-shifted particles as $s = 0.48$ and $s = 0.6$, respectively, (Sec.~\ref{sec:results_magnetic_chain_formation}) and the correct corner particle orientation in 12-particle magic clusters of radially-shifted particles (Sec.~\ref{sec:results_frustrated}). 
It is also able to reproduce the trend in the order parameters $S^{(e)}$ and $\langle \boldsymbol{n} \cdot \boldsymbol{h} \rangle$ as functions of field strength (Sec.~\ref{sec:results_lateral_field}) and the nucleation phase for the self-assembly of laterally-shifted particles (Sec.~\ref{sec:results_nucleation_factor}). 
While making accurate particle orientation predictions, phenomena depending on cluster diffusion, multi-particle translation, and cluster reorientation are less well represented, as clearly shown, for example, by the $\theta_{nn} = 0$ peak for $s \neq 0$ (Sec.~\ref{sec:results_magnetic_chain_formation}), scaling factor in Sec.~\ref{sec:results_lateral_field}, and the lack of a growth phase (Sec.~\ref{sec:results_nucleation_factor}).  
In the DE simulation used here, each generation of chromosomes must be either as or more energetically favorable than the last.
Therefore, the DE simulation cannot easily move formed clusters, as it has to first disassemble them, making energetically unfavorable transition chromosomes. 
Further, reorienting particles inside a cluster is also difficult because changing one particle destabilizes the interactions with its nearest neighbors that cannot rotate out-of-plane to make minor adjustments to compensate. 

Modifications to the next generation of DE simulation should include cluster-level moves, local relaxation, and continuous out-of-plane rotation.
Including cluster-level moves that allow fully-formed structures to diffuse and combine will address the "freezing" and missing growth phase. 
Cluster-level moves could be in part implemented by using the genetic crossover found in genetic algorithms.~\cite{katoch2021b}
Local optimization~\cite{bianchi2012} and additional simulated annealing,~\cite{biswas2017} which has been combined with genetic algorithms in the past, and periodic relaxing of the fitness function will increase ability for the simulation to explore less-fit chromosomes.
Out-of-plane rotation will allow subtle adjustments within formed clusters with less energetic penalty allowing for finer adjustments of chain and cluster structures.

This work serves as a proof-of-concept for new possibilities for DE-based simulations to investigate different types of particle-particle interactions by modifying the fitness function to include different patch shapes and materials.
The fitness function can be expanded to include electric-field and multiple patchy interactions.
The DE parameters ($c_F$, $c_{mut}$, and $c_R$) can be adjusted to improve the convergence time of the simulations.
In addition, DE simulation predictions can be coupled with other modeling methods that will allow forecasting of macroscopic properties for MR fluids.

\begin{acknowledgments}
EM and IK acknowledge the support from the NSF Phase II CREST Center for Interface Design and Engineered Assembly of Low Dimensional Materials (IDEALS), NSF grant number EES-2112550.
\end{acknowledgments}
\section*{Author Declarations}
\subsection*{Conflict of Interest}
The authors have no conflict of interest to disclose.

\subsection*{Author Contributions}
E.M., K.K., and I.K. discussed and conceptualized the project. 
E.M. developed the code with feedback from K.K., ran the simulations, and provided a first draft of the manuscript. 
E.M., K.K., and I.K. discussed the interpretation of the simulation results in context of the case studies. 
K.K., and I.K. provided feedback on the manuscript. 
All authors have read and approved the final version.

\section*{Data Availability Statement}

The python package developed and the data that support the findings of this study are openly available at the GitHub repository found at \href{https://github.com/emcpher001/Janus-Differential-Evolution}{https://github.com/emcpher001/Janus-Differential-Evolution}.

\clearpage

\section{Supplementary Material}

\renewcommand{\figurename}{Figure}\renewcommand{\thefigure}{S\arabic{figure}} 
\renewcommand{\thepage}{S\arabic{page}} 
\renewcommand{\thesection}{S\arabic{section}}  
\renewcommand{\thetable}{S\arabic{table}}  
\renewcommand{\theequation}{S\arabic{equation}}
\setcounter{figure}{0}
\setcounter{section}{0}

The supporting information document provides the simulation parameters (Sec.~\ref{sec:parameters}), the system box size calculation (Sec.~\ref{sec:box}), information on the impact of system size on number of generations needed for convergence (Sec.~\ref{sec:size}), Kernel density estimates for different particle sizes (Sec.~\ref{sec:radius}) and intermediate shifts between $s = 0.4-0.6$ (Sec.~\ref{sec:focus}), comparison of $\theta_{nn}$ values from DE simulation with theoretical limits (Sec.~\ref{sec:energy}), and configuration data for 12-particle frustrated clusters at $s = 0.0$ and $0.58$ (Sec.~\ref{sec:frustrated}), and DE simulation results for the changing field strength case study using $c_F=0.5$ (Sec.~\ref{sec:F}).

\section{Simulation Physical Parameters}\label{sec:parameters}

\begin{table}[!h]
\begin{tabular}{llll}
\textbf{Case Study} & \textbf{Particle Radius (\SI{}{m})} & \textbf{Magnetic Dipole Strength (\SI{}{Am^2})} & \textbf{Volume Fraction} \\ \hline
A                   & \SI{1.2e-6}{}                       & \SI{5e-14}{}                                                     & 0.01                     \\
B                   & \SI{2.375e-6}{}                     & \SI{2.8e-15}{}                                                 & N/A                      \\
C                   & \SI{2e-6}{}                         & \SI{5e-14}{}                                                     & 0.1                      \\
D                   & \SI{2e-6}{}                         & \SI{5e-15}{}                                                     & 0.007                   
\end{tabular}
\caption{\label{tab:parameters}The physical parameters used for each simulation.}
\end{table}

Table~\ref{tab:parameters} shows the physical parameters used for the DE simulation for each case study.
The case study letter corresponds with the model systems subsections in the article. 
Where possible, physical parameters from the relevant case study were used.

\section{Box Size Determination}\label{sec:box}
The box size, $L$, is calculated using a volumetric fraction $v_{frac}$, and assuming a volume that is one particle thick in the out-of-plane direction, using Eqn.~\ref{eqn:boxsize}.
\begin{equation}
	\label{eqn:boxsize}
	L = \sqrt{\frac{\pi N}{24\cdot v_{frac}}}  
\end{equation}
The chromosome distance is converted to real-space location by multiplying the gene by L.

\clearpage

\section{Impact of System Size} \label{sec:size}

\begin{figure}[h!]
    \centering
	\makebox[0pt]{\includegraphics[width=4.5in]{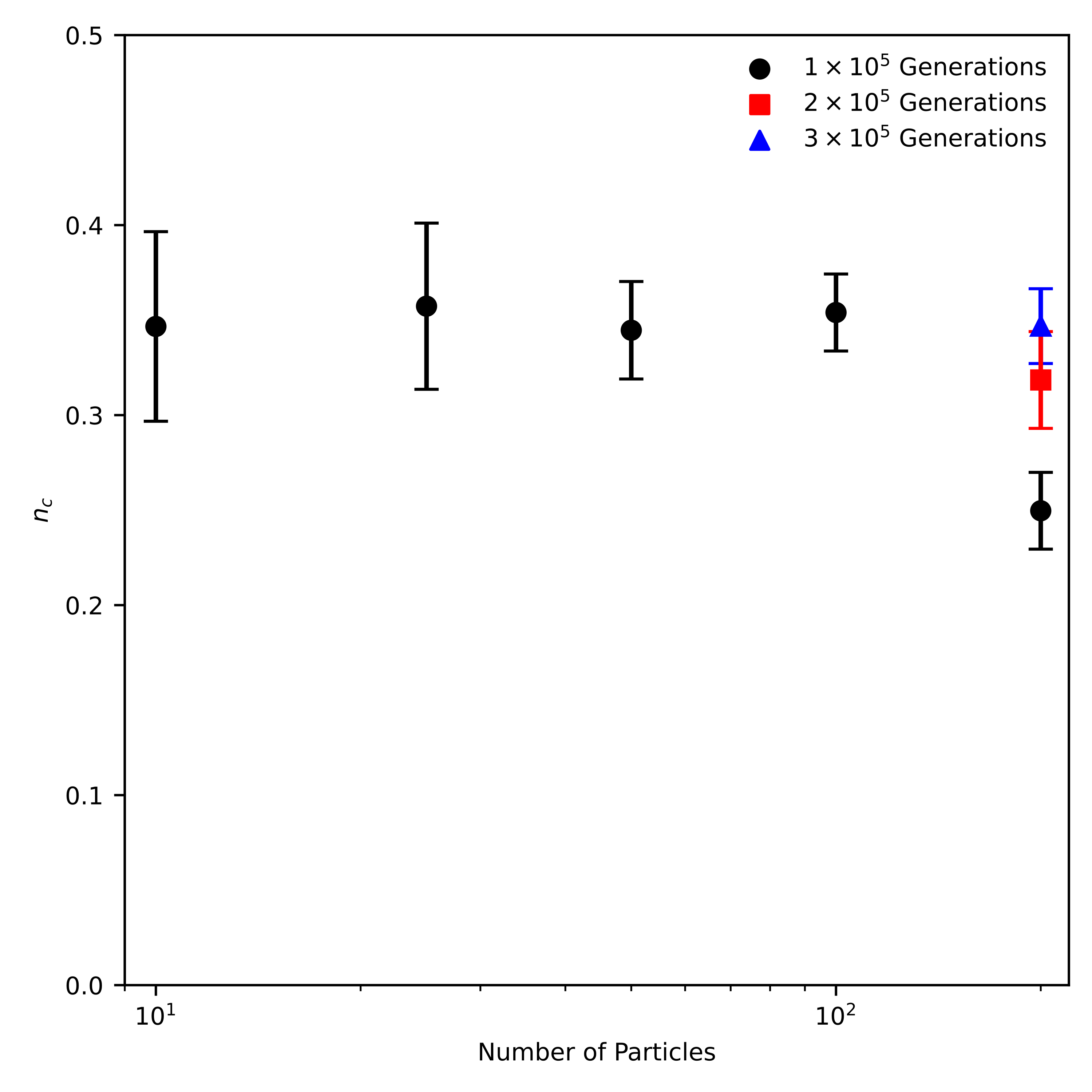}}
	\caption{\label{fig:SI_particleCount} Plot showing the average nucleation factor, $n_c$, for 30 DE simulations for different numbers of Janus particles, where error bars show one standard deviation.
    Above 100 particles, more generations are needed for the optimization to converge. 
    However, when run for more generations, higher particle counts show a minimal change in predicted $n_c$ and the variance in the predictions remains steady, indicating that simulating more than 100 particles does not affect or improve the predictions, but substantially increases the computational cost as the number of generations increases.
    }
\end{figure}

\pagebreak

\section{Kernel Density Estimates Showing the Effect of Particle Size}\label{sec:radius}

\begin{figure}[h!]
    \centering
	\makebox[0pt]{\includegraphics[width=6.5in]{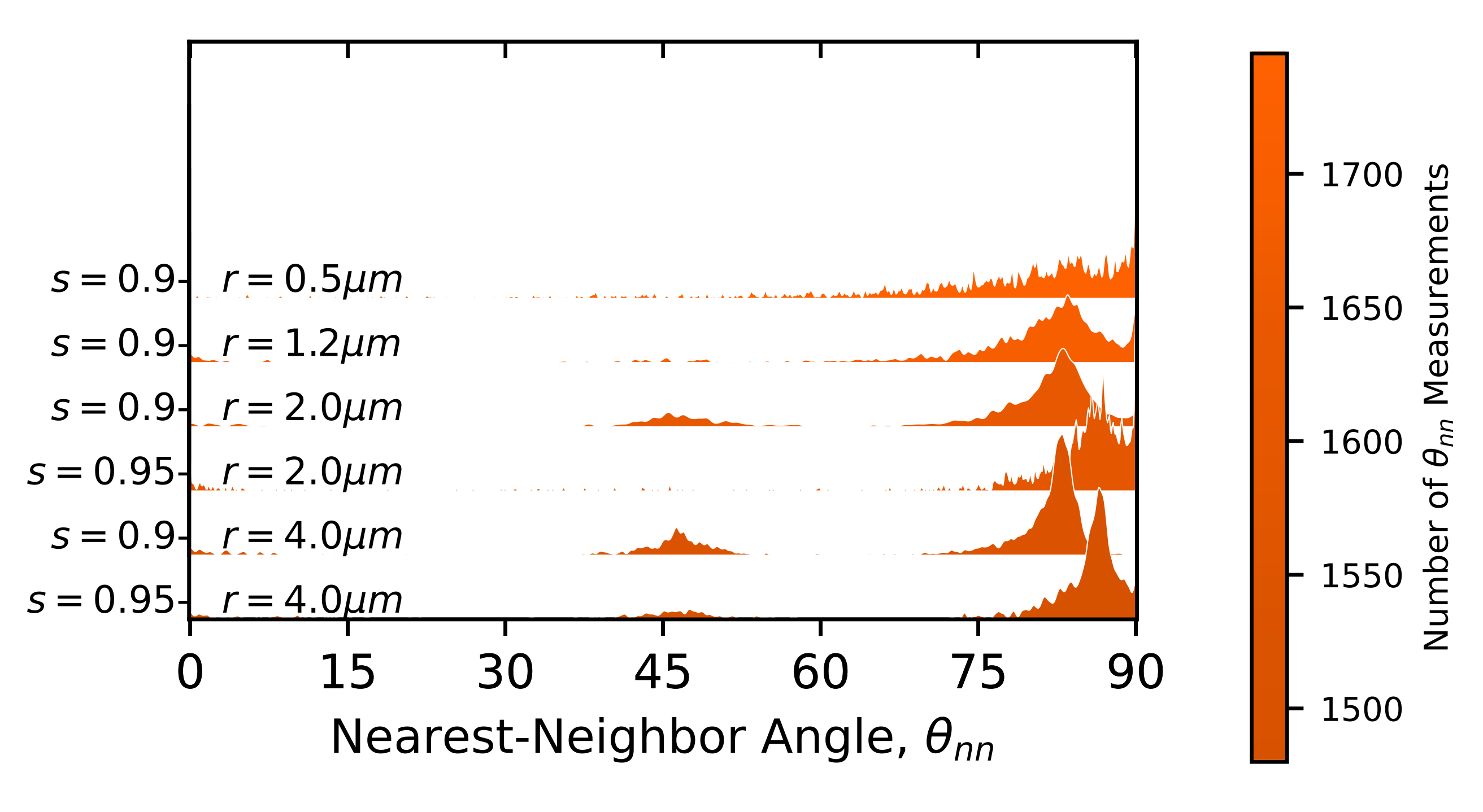}}
	\caption{\label{fig:particleSize} 
    Kernel density estimate of $\theta_{nn}$ as a function of both shift ($s=0.9$ and $0.95$) and particle radius ($r=$\SI{0.5} to \SI{4}{\mu m}) for DE simulations.
    For a given $s$, changing $r$ does not affect the location of the peak near $\theta_{nn} \approx 80\degree$ corresponding to doublets.
    The larger the radius, the more pronounced chain formation is, i.e., a small peak is observed at $\theta_{nn} \approx 45\degree$.
    The radius dependence is explained by the DE simulation using a non-renormalized $\lambda$, as discussed in the main manuscript. 
}

\end{figure}

\pagebreak

\section{Kernel Density Estimate for Intermediate Shifts in Range s = 0.4-0.6} \label{sec:focus}

\begin{figure}[h!]
    \centering
	\makebox[0pt]{\includegraphics[width=6.5in]{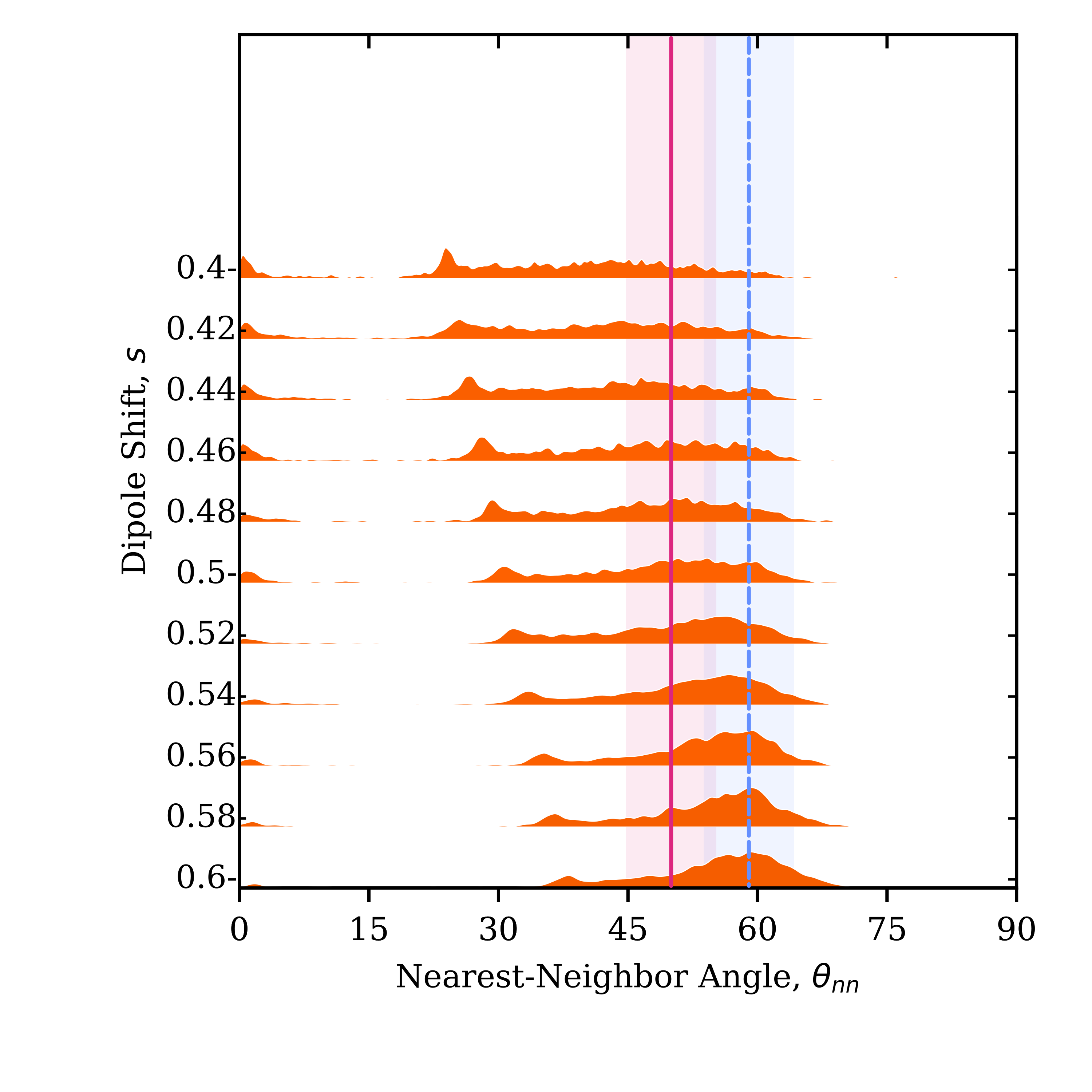}}
	\caption{\label{fig:SI_shiftFocus} Kernel density estimate of $\theta_{nn}$ obtained from simulations run between $s=0.4$ and $s=0.6$ in $=0.02$ increments.
    The KDEs show that the $\theta_{nn}$ distribution stays broad from $s=0.4-0.54$. It seems to peak at 0.56 and then tightens to the more clear peak at $s=0.6$.
}
\end{figure}

\pagebreak

\section{Comparison of DE Prediction for $\theta_{nm}$ with Theoretical Limit} \label{sec:energy}

\begin{figure}[h!]
    \centering
	\makebox[0pt]{\includegraphics[width=6.5in]{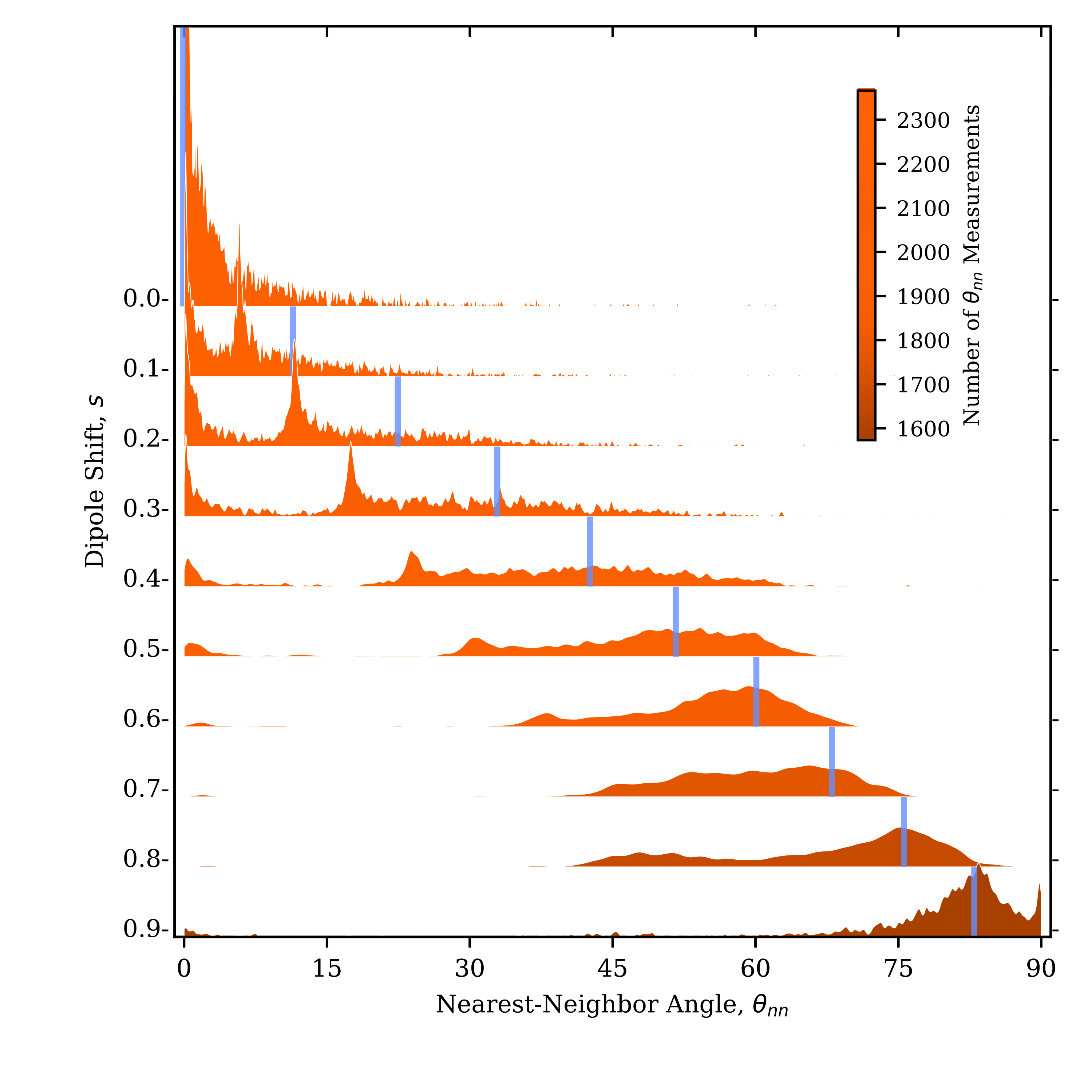}}
	\caption{\label{fig:SI_shiftsMinimumDoubletEnergy}
    The same kernel density estimates of $\theta_{nn}$ for different $s$ as shown in Fig. 3 in the manuscript, but with the calculated dipole-dipole energy minima for a pair of particles overlaid (vertical blue lines).
    The energy minima are calculated using Eq. 7, assuming both particle dipoles point the same direction as the external field.
    The DE predictions have peaks close to the calculated minima for $s=0.0$ and $s=0.6-0.9$, but deviate for $s=0.1-0.5$. 
}

\end{figure}

\pagebreak

\section{12-Particle Frustrated Cluster Structure Data for s = 0.0 and s = 0.58} \label{sec:frustrated}

\begin{figure}[h!]
    \centering
	\makebox[0pt]{\includegraphics[width=6.2in]{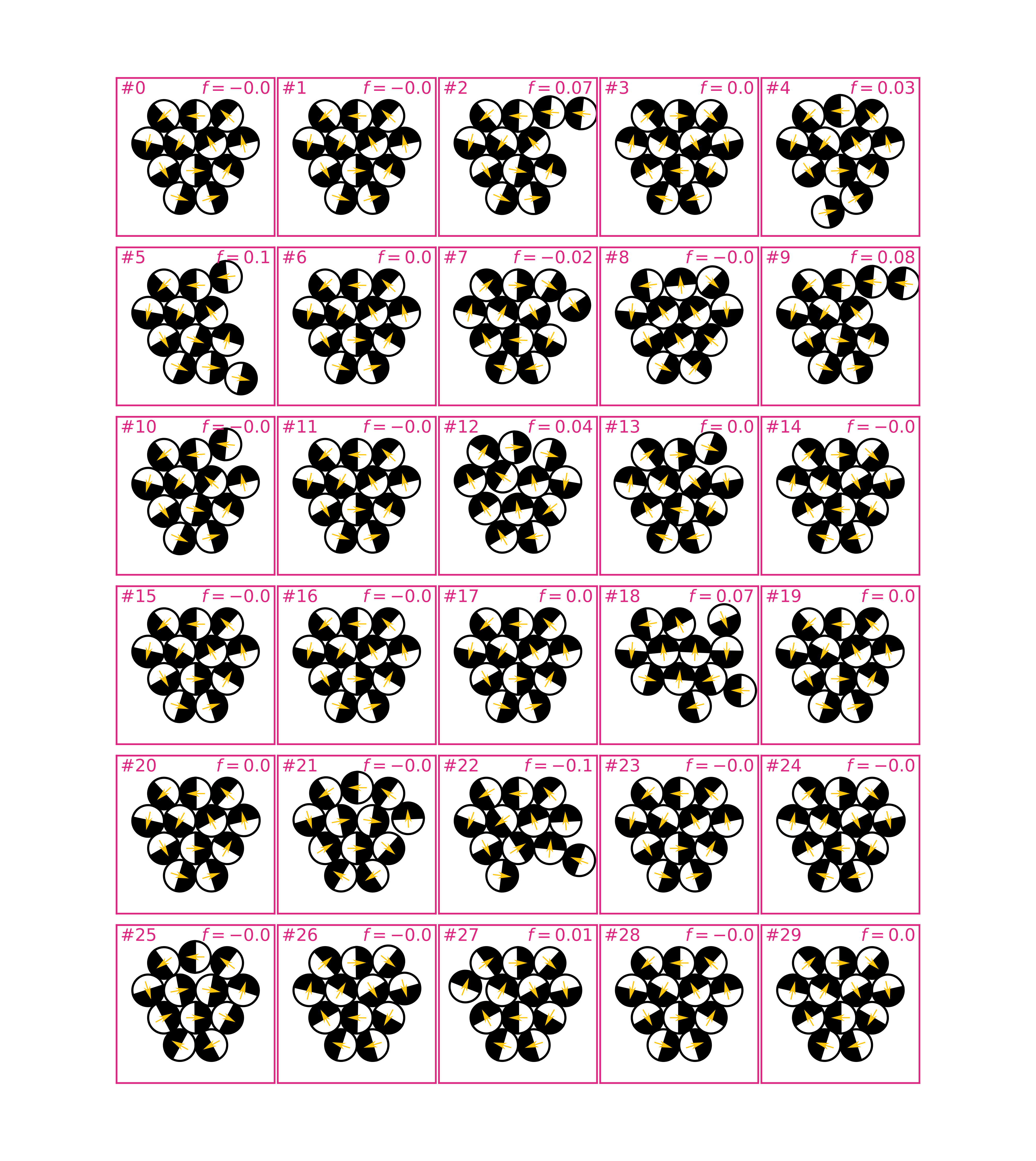}}
	\caption{\label{fig:SI_allFrustrated_s0.0}
    The final, optimized structures for each of the 30 DE simulations discussed in Section 4b for $s=0.0$.
    The boxes and text associated with structures with a similarity score of $f < 0.95$ are highlighted in red.
    Note that some structures have a tendency to break apart, and most form head-to-tail loop-like structures.
}
\end{figure}

\begin{figure}[h!]
    \centering
	\makebox[0pt]{\includegraphics[width=6.2in]{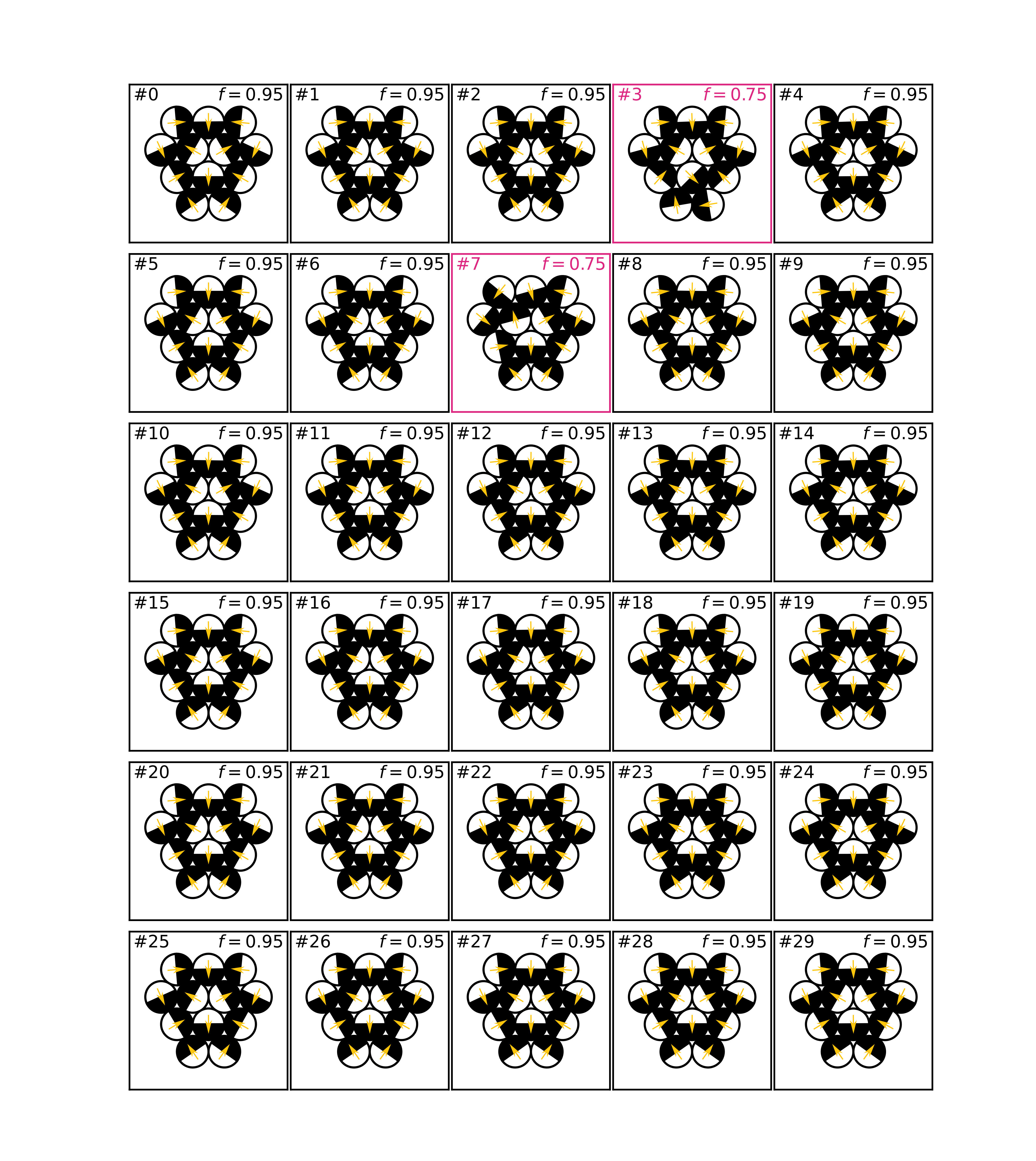}}
	\caption{\label{fig:SI_allFrustrated_s0.58}
    The final, optimized structures for each of the 30 DE simulations discussed in Section 4b for $s=0.58$.
    The boxes and text associated with structures with a similarity score of $f < 0.95$ are highlighted in red.
}
\end{figure}

\clearpage
\section{Mutation factor $F$ and Changing Field Strength}\label{sec:F}

\begin{figure}[h!]
    \centering
	\makebox[0pt]{\includegraphics[width=6.2in]{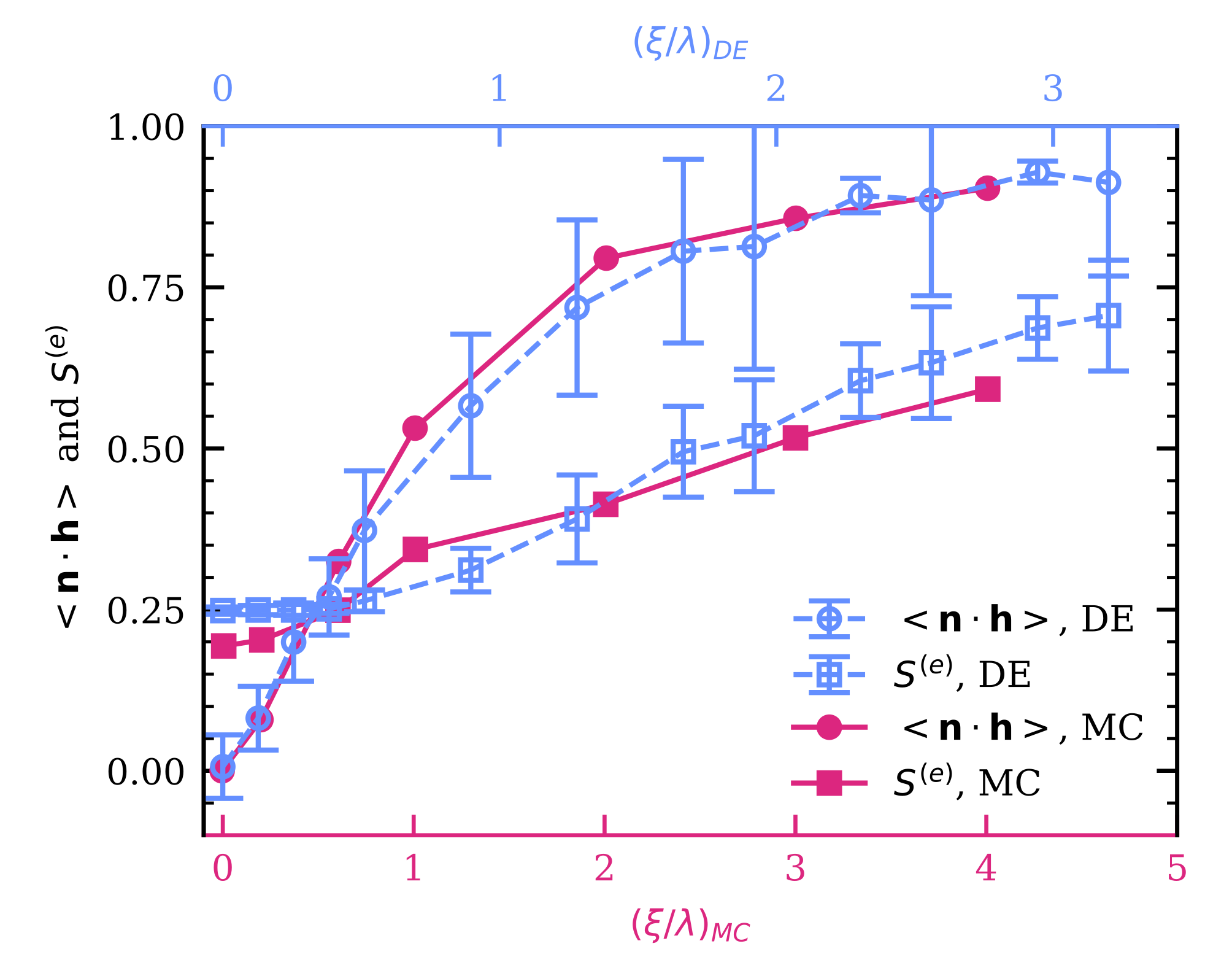}}
	\caption{\label{fig:0.5F_structure} 
    Change in order parameters $S^{(e)}$ (squares) and $\langle \boldsymbol{n} \cdot \boldsymbol{h} \rangle$ (circles) as a function of external magnetic field at constant $ \lambda = 5$. 
    DE simulation results were run with $c_F = 0.5$, and the scaling between the top, blue and bottom, red axes is 1.45.
    Note the larger standard deviations for the DE results compared to Fig. 5 in the main manuscript. 
    Parameters are plotted against $(\xi / \lambda)_{MC}$ for the MC simulation data from Okada and Satoh, see Ref. 40 in the main manuscript (red, solid line and symbols), and $(\xi / \lambda)_{DE}$ for the DE simulation (blue, dashed line and open symbols).
}
\end{figure}

\bibliography{DE_paper}

\begin{thebibliography}{51}%
\makeatletter
\providecommand \@ifxundefined [1]{%
 \@ifx{#1\undefined}
}%
\providecommand \@ifnum [1]{%
 \ifnum #1\expandafter \@firstoftwo
 \else \expandafter \@secondoftwo
 \fi
}%
\providecommand \@ifx [1]{%
 \ifx #1\expandafter \@firstoftwo
 \else \expandafter \@secondoftwo
 \fi
}%
\providecommand \natexlab [1]{#1}%
\providecommand \enquote  [1]{``#1''}%
\providecommand \bibnamefont  [1]{#1}%
\providecommand \bibfnamefont [1]{#1}%
\providecommand \citenamefont [1]{#1}%
\providecommand \href@noop [0]{\@secondoftwo}%
\providecommand \href [0]{\begingroup \@sanitize@url \@href}%
\providecommand \@href[1]{\@@startlink{#1}\@@href}%
\providecommand \@@href[1]{\endgroup#1\@@endlink}%
\providecommand \@sanitize@url [0]{\catcode `\\12\catcode `\$12\catcode
  `\&12\catcode `\#12\catcode `\^12\catcode `\_12\catcode `\%12\relax}%
\providecommand \@@startlink[1]{}%
\providecommand \@@endlink[0]{}%
\providecommand \url  [0]{\begingroup\@sanitize@url \@url }%
\providecommand \@url [1]{\endgroup\@href {#1}{\urlprefix }}%
\providecommand \urlprefix  [0]{URL }%
\providecommand \Eprint [0]{\href }%
\providecommand \doibase [0]{https://doi.org/}%
\providecommand \selectlanguage [0]{\@gobble}%
\providecommand \bibinfo  [0]{\@secondoftwo}%
\providecommand \bibfield  [0]{\@secondoftwo}%
\providecommand \translation [1]{[#1]}%
\providecommand \BibitemOpen [0]{}%
\providecommand \bibitemStop [0]{}%
\providecommand \bibitemNoStop [0]{.\EOS\space}%
\providecommand \EOS [0]{\spacefactor3000\relax}%
\providecommand \BibitemShut  [1]{\csname bibitem#1\endcsname}%
\let\auto@bib@innerbib\@empty
\bibitem [{\citenamefont {Ashtiani}, \citenamefont {Hashemabadi},\ and\
  \citenamefont {Ghaffari}(2015)}]{ashtiani2015}%
  \BibitemOpen
  \bibfield  {author} {\bibinfo {author} {\bibfnamefont {M.}~\bibnamefont
  {Ashtiani}}, \bibinfo {author} {\bibfnamefont {S.~H.}\ \bibnamefont
  {Hashemabadi}},\ and\ \bibinfo {author} {\bibfnamefont {A.}~\bibnamefont
  {Ghaffari}},\ }\bibfield  {title} {\enquote {\bibinfo {title} {A review on
  the magnetorheological fluid preparation and stabilization},}\ }\href
  {https://doi.org/10.1016/j.jmmm.2014.09.020} {\bibfield  {journal} {\bibinfo
  {journal} {Journal of Magnetism and Magnetic Materials}\ }\textbf {\bibinfo
  {volume} {374}},\ \bibinfo {pages} {716--730} (\bibinfo {year}
  {2015})}\BibitemShut {NoStop}%
\bibitem [{\citenamefont {Eshgarf}, \citenamefont {Ahmadi~Nadooshan},\ and\
  \citenamefont {Raisi}(2022)}]{eshgarf2022}%
  \BibitemOpen
  \bibfield  {author} {\bibinfo {author} {\bibfnamefont {H.}~\bibnamefont
  {Eshgarf}}, \bibinfo {author} {\bibfnamefont {A.}~\bibnamefont
  {Ahmadi~Nadooshan}},\ and\ \bibinfo {author} {\bibfnamefont {A.}~\bibnamefont
  {Raisi}},\ }\bibfield  {title} {\enquote {\bibinfo {title} {An overview on
  properties and applications of magnetorheological fluids: {{Dampers}},
  batteries, valves and brakes},}\ }\href
  {https://doi.org/10.1016/j.est.2022.104648} {\bibfield  {journal} {\bibinfo
  {journal} {Journal of Energy Storage}\ }\textbf {\bibinfo {volume} {50}},\
  \bibinfo {pages} {104648} (\bibinfo {year} {2022})}\BibitemShut {NoStop}%
\bibitem [{\citenamefont {Olabi}\ and\ \citenamefont
  {Grunwald}(2007)}]{olabi2007}%
  \BibitemOpen
  \bibfield  {author} {\bibinfo {author} {\bibfnamefont {A.~G.}\ \bibnamefont
  {Olabi}}\ and\ \bibinfo {author} {\bibfnamefont {A.}~\bibnamefont
  {Grunwald}},\ }\bibfield  {title} {\enquote {\bibinfo {title} {Design and
  application of magneto-rheological fluid},}\ }\href
  {https://doi.org/10.1016/j.matdes.2006.10.009} {\bibfield  {journal}
  {\bibinfo  {journal} {Materials \& Design}\ }\textbf {\bibinfo {volume}
  {28}},\ \bibinfo {pages} {2658--2664} (\bibinfo {year} {2007})}\BibitemShut
  {NoStop}%
\bibitem [{\citenamefont {Jonsdottir}\ \emph {et~al.}(2009)\citenamefont
  {Jonsdottir}, \citenamefont {Thorarinsson}, \citenamefont {Palsson},\ and\
  \citenamefont {Gudmundsson}}]{jonsdottir2009}%
  \BibitemOpen
  \bibfield  {author} {\bibinfo {author} {\bibfnamefont {F.}~\bibnamefont
  {Jonsdottir}}, \bibinfo {author} {\bibfnamefont {E.}~\bibnamefont
  {Thorarinsson}}, \bibinfo {author} {\bibfnamefont {H.}~\bibnamefont
  {Palsson}},\ and\ \bibinfo {author} {\bibfnamefont {K.}~\bibnamefont
  {Gudmundsson}},\ }\bibfield  {title} {\enquote {\bibinfo {title} {Influence
  of {{Parameter Variations}} on the {{Braking Torque}} of a
  {{Magnetorheological Prosthetic Knee}}},}\ }\href
  {https://doi.org/10.1177/1045389X08094303} {\bibfield  {journal} {\bibinfo
  {journal} {Journal of Intelligent Material Systems and Structures}\ }\textbf
  {\bibinfo {volume} {20}},\ \bibinfo {pages} {659--667} (\bibinfo {year}
  {2009})}\BibitemShut {NoStop}%
\bibitem [{\citenamefont {Karakoc}, \citenamefont {Park},\ and\ \citenamefont
  {Suleman}(2008)}]{karakoc2008}%
  \BibitemOpen
  \bibfield  {author} {\bibinfo {author} {\bibfnamefont {K.}~\bibnamefont
  {Karakoc}}, \bibinfo {author} {\bibfnamefont {E.~J.}\ \bibnamefont {Park}},\
  and\ \bibinfo {author} {\bibfnamefont {A.}~\bibnamefont {Suleman}},\
  }\bibfield  {title} {\enquote {\bibinfo {title} {Design considerations for an
  automotive magnetorheological brake},}\ }\href
  {https://doi.org/10.1016/j.mechatronics.2008.02.003} {\bibfield  {journal}
  {\bibinfo  {journal} {Mechatronics}\ }\textbf {\bibinfo {volume} {18}},\
  \bibinfo {pages} {434--447} (\bibinfo {year} {2008})}\BibitemShut {NoStop}%
\bibitem [{\citenamefont {Li}, \citenamefont {Yang},\ and\ \citenamefont
  {Yin}(2020)}]{li2020a}%
  \BibitemOpen
  \bibfield  {author} {\bibinfo {author} {\bibfnamefont {Z.}~\bibnamefont
  {Li}}, \bibinfo {author} {\bibfnamefont {F.}~\bibnamefont {Yang}},\ and\
  \bibinfo {author} {\bibfnamefont {Y.}~\bibnamefont {Yin}},\ }\bibfield
  {title} {\enquote {\bibinfo {title} {Smart {{Materials}} by {{Nanoscale
  Magnetic Assembly}}},}\ }\href {https://doi.org/10.1002/adfm.201903467}
  {\bibfield  {journal} {\bibinfo  {journal} {Advanced Functional Materials}\
  }\textbf {\bibinfo {volume} {30}},\ \bibinfo {pages} {1903467} (\bibinfo
  {year} {2020})}\BibitemShut {NoStop}%
\bibitem [{\citenamefont {Kumar}\ \emph {et~al.}(2022)\citenamefont {Kumar},
  \citenamefont {Kumar}, \citenamefont {Bharti}, \citenamefont {Yadav},\ and\
  \citenamefont {Das}}]{kumar2022}%
  \BibitemOpen
  \bibfield  {author} {\bibinfo {author} {\bibfnamefont {M.}~\bibnamefont
  {Kumar}}, \bibinfo {author} {\bibfnamefont {A.}~\bibnamefont {Kumar}},
  \bibinfo {author} {\bibfnamefont {R.~K.}\ \bibnamefont {Bharti}}, \bibinfo
  {author} {\bibfnamefont {H.~N.~S.}\ \bibnamefont {Yadav}},\ and\ \bibinfo
  {author} {\bibfnamefont {M.}~\bibnamefont {Das}},\ }\bibfield  {title}
  {\enquote {\bibinfo {title} {A review on rheological properties of
  magnetorheological fluid for engineering components polishing},}\ }\href
  {https://doi.org/10.1016/j.matpr.2021.11.611} {\bibfield  {journal} {\bibinfo
   {journal} {Materials Today: Proceedings}\ }\bibinfo {series} {First
  {{International Conference}} on {{Advances}} in {{Mechanical Engineering}}
  and {{Material Science}}},\ \textbf {\bibinfo {volume} {56}},\ \bibinfo
  {pages} {A6--A12} (\bibinfo {year} {2022})}\BibitemShut {NoStop}%
\bibitem [{\citenamefont {Chen}\ \emph {et~al.}(2012)\citenamefont {Chen},
  \citenamefont {Yan}, \citenamefont {Zhang}, \citenamefont {Bae},\ and\
  \citenamefont {Granick}}]{chen2012}%
  \BibitemOpen
  \bibfield  {author} {\bibinfo {author} {\bibfnamefont {Q.}~\bibnamefont
  {Chen}}, \bibinfo {author} {\bibfnamefont {J.}~\bibnamefont {Yan}}, \bibinfo
  {author} {\bibfnamefont {J.}~\bibnamefont {Zhang}}, \bibinfo {author}
  {\bibfnamefont {S.~C.}\ \bibnamefont {Bae}},\ and\ \bibinfo {author}
  {\bibfnamefont {S.}~\bibnamefont {Granick}},\ }\bibfield  {title} {\enquote
  {\bibinfo {title} {Janus and {{Multiblock Colloidal Particles}}},}\ }\href
  {https://doi.org/10.1021/la302226w} {\bibfield  {journal} {\bibinfo
  {journal} {Langmuir}\ }\textbf {\bibinfo {volume} {28}},\ \bibinfo {pages}
  {13555--13561} (\bibinfo {year} {2012})}\BibitemShut {NoStop}%
\bibitem [{\citenamefont {Zhang}, \citenamefont {Luijten},\ and\ \citenamefont
  {Granick}(2015)}]{zhang2015}%
  \BibitemOpen
  \bibfield  {author} {\bibinfo {author} {\bibfnamefont {J.}~\bibnamefont
  {Zhang}}, \bibinfo {author} {\bibfnamefont {E.}~\bibnamefont {Luijten}},\
  and\ \bibinfo {author} {\bibfnamefont {S.}~\bibnamefont {Granick}},\
  }\bibfield  {title} {\enquote {\bibinfo {title} {Toward {{Design Rules}} of
  {{Directional Janus Colloidal Assembly}}},}\ }\href
  {https://doi.org/10.1146/annurev-physchem-040214-121241} {\bibfield
  {journal} {\bibinfo  {journal} {Annual Review of Physical Chemistry, Vol 66}\
  }\textbf {\bibinfo {volume} {66}},\ \bibinfo {pages} {581--600} (\bibinfo
  {year} {2015})}\BibitemShut {NoStop}%
\bibitem [{\citenamefont {Zhang}, \citenamefont {Grzybowski},\ and\
  \citenamefont {Granick}(2017)}]{zhang2017}%
  \BibitemOpen
  \bibfield  {author} {\bibinfo {author} {\bibfnamefont {J.}~\bibnamefont
  {Zhang}}, \bibinfo {author} {\bibfnamefont {B.~A.}\ \bibnamefont
  {Grzybowski}},\ and\ \bibinfo {author} {\bibfnamefont {S.}~\bibnamefont
  {Granick}},\ }\bibfield  {title} {\enquote {\bibinfo {title} {Janus
  {{Particle Synthesis}}, {{Assembly}}, and {{Application}}},}\ }\href
  {https://doi.org/10.1021/acs.langmuir.7b01123} {\bibfield  {journal}
  {\bibinfo  {journal} {Langmuir}\ }\textbf {\bibinfo {volume} {33}},\ \bibinfo
  {pages} {6964--6977} (\bibinfo {year} {2017})}\BibitemShut {NoStop}%
\bibitem [{\citenamefont {Bishop}, \citenamefont {Biswal},\ and\ \citenamefont
  {Bharti}(2023)}]{bishop2023}%
  \BibitemOpen
  \bibfield  {author} {\bibinfo {author} {\bibfnamefont {K.~J.}\ \bibnamefont
  {Bishop}}, \bibinfo {author} {\bibfnamefont {S.~L.}\ \bibnamefont {Biswal}},\
  and\ \bibinfo {author} {\bibfnamefont {B.}~\bibnamefont {Bharti}},\
  }\bibfield  {title} {\enquote {\bibinfo {title} {Active {{Colloids}} as
  {{Models}}, {{Materials}}, and {{Machines}}},}\ }\href
  {https://doi.org/10.1146/annurev-chembioeng-101121-084939} {\bibfield
  {journal} {\bibinfo  {journal} {Annual Review of Chemical and Biomolecular
  Engineering}\ }\textbf {\bibinfo {volume} {14}},\ \bibinfo {pages} {1--30}
  (\bibinfo {year} {2023})}\BibitemShut {NoStop}%
\bibitem [{\citenamefont {Smoukov}\ \emph {et~al.}(2009)\citenamefont
  {Smoukov}, \citenamefont {Gangwal}, \citenamefont {Marquez},\ and\
  \citenamefont {Velev}}]{smoukov2009b}%
  \BibitemOpen
  \bibfield  {author} {\bibinfo {author} {\bibfnamefont {S.~K.}\ \bibnamefont
  {Smoukov}}, \bibinfo {author} {\bibfnamefont {S.}~\bibnamefont {Gangwal}},
  \bibinfo {author} {\bibfnamefont {M.}~\bibnamefont {Marquez}},\ and\ \bibinfo
  {author} {\bibfnamefont {O.~D.}\ \bibnamefont {Velev}},\ }\bibfield  {title}
  {\enquote {\bibinfo {title} {Reconfigurable responsive structures assembled
  from magnetic {{Janus}} particles},}\ }\href
  {https://doi.org/10.1039/b814304h} {\bibfield  {journal} {\bibinfo  {journal}
  {Soft Matter}\ }\textbf {\bibinfo {volume} {5}},\ \bibinfo {pages} {1285}
  (\bibinfo {year} {2009})}\BibitemShut {NoStop}%
\bibitem [{\citenamefont {Ren}\ \emph {et~al.}(2012)\citenamefont {Ren},
  \citenamefont {Ruditskiy}, \citenamefont {Song},\ and\ \citenamefont
  {Kretzschmar}}]{ren2012}%
  \BibitemOpen
  \bibfield  {author} {\bibinfo {author} {\bibfnamefont {B.}~\bibnamefont
  {Ren}}, \bibinfo {author} {\bibfnamefont {A.}~\bibnamefont {Ruditskiy}},
  \bibinfo {author} {\bibfnamefont {J.~H.~K.}\ \bibnamefont {Song}},\ and\
  \bibinfo {author} {\bibfnamefont {I.}~\bibnamefont {Kretzschmar}},\
  }\bibfield  {title} {\enquote {\bibinfo {title} {Assembly {{Behavior}} of
  {{Iron Oxide-Capped Janus Particles}} in a {{Magnetic Field}}},}\ }\href
  {https://doi.org/10.1021/la203969f} {\bibfield  {journal} {\bibinfo
  {journal} {Langmuir}\ }\textbf {\bibinfo {volume} {28}},\ \bibinfo {pages}
  {1149--1156} (\bibinfo {year} {2012})}\BibitemShut {NoStop}%
\bibitem [{\citenamefont {Sinn}\ \emph {et~al.}(2011)\citenamefont {Sinn},
  \citenamefont {Kinnunen}, \citenamefont {Pei}, \citenamefont {Clarke},
  \citenamefont {McNaughton},\ and\ \citenamefont {Kopelman}}]{sinn2011}%
  \BibitemOpen
  \bibfield  {author} {\bibinfo {author} {\bibfnamefont {I.}~\bibnamefont
  {Sinn}}, \bibinfo {author} {\bibfnamefont {P.}~\bibnamefont {Kinnunen}},
  \bibinfo {author} {\bibfnamefont {S.~N.}\ \bibnamefont {Pei}}, \bibinfo
  {author} {\bibfnamefont {R.}~\bibnamefont {Clarke}}, \bibinfo {author}
  {\bibfnamefont {B.~H.}\ \bibnamefont {McNaughton}},\ and\ \bibinfo {author}
  {\bibfnamefont {R.}~\bibnamefont {Kopelman}},\ }\bibfield  {title} {\enquote
  {\bibinfo {title} {Magnetically uniform and tunable {{Janus}} particles},}\
  }\href {https://doi.org/10.1063/1.3541876} {\bibfield  {journal} {\bibinfo
  {journal} {Applied Physics Letters}\ }\textbf {\bibinfo {volume} {98}},\
  \bibinfo {pages} {024101} (\bibinfo {year} {2011})}\BibitemShut {NoStop}%
\bibitem [{\citenamefont {Demir{\"o}rs}\ \emph {et~al.}(2018)\citenamefont
  {Demir{\"o}rs}, \citenamefont {Akan}, \citenamefont {Poloni},\ and\
  \citenamefont {Studart}}]{demirors2018a}%
  \BibitemOpen
  \bibfield  {author} {\bibinfo {author} {\bibfnamefont {A.~F.}\ \bibnamefont
  {Demir{\"o}rs}}, \bibinfo {author} {\bibfnamefont {M.~T.}\ \bibnamefont
  {Akan}}, \bibinfo {author} {\bibfnamefont {E.}~\bibnamefont {Poloni}},\ and\
  \bibinfo {author} {\bibfnamefont {A.~R.}\ \bibnamefont {Studart}},\
  }\bibfield  {title} {\enquote {\bibinfo {title} {Active cargo transport with
  {{Janus}} colloidal shuttles using electric and magnetic fields},}\ }\href
  {https://doi.org/10.1039/C8SM00513C} {\bibfield  {journal} {\bibinfo
  {journal} {Soft Matter}\ }\textbf {\bibinfo {volume} {14}},\ \bibinfo {pages}
  {4741--4749} (\bibinfo {year} {2018})}\BibitemShut {NoStop}%
\bibitem [{\citenamefont {Suzuki}(1995)}]{suzuki1995}%
  \BibitemOpen
  \bibfield  {author} {\bibinfo {author} {\bibfnamefont {T.}~\bibnamefont
  {Suzuki}},\ }\bibfield  {title} {\enquote {\bibinfo {title} {Coercivity
  mechanism in ({{CoPt}}) and ({{CoPd}}) multilayers},}\ }\href
  {https://doi.org/10.1016/0956-716X(95)00519-2} {\bibfield  {journal}
  {\bibinfo  {journal} {Scripta Metallurgica et Materialia}\ }\bibinfo {series}
  {Proceedings of an {{Acta Metallurgica Meeting}} on {{Novel Magnetic
  Structures}} and {{Properties}}},\ \textbf {\bibinfo {volume} {33}},\
  \bibinfo {pages} {1609--1623} (\bibinfo {year} {1995})}\BibitemShut {NoStop}%
\bibitem [{\citenamefont {Baraban}\ \emph {et~al.}(2008)\citenamefont
  {Baraban}, \citenamefont {Makarov}, \citenamefont {Albrecht}, \citenamefont
  {Rivier}, \citenamefont {Leiderer},\ and\ \citenamefont
  {Erbe}}]{baraban2008}%
  \BibitemOpen
  \bibfield  {author} {\bibinfo {author} {\bibfnamefont {L.}~\bibnamefont
  {Baraban}}, \bibinfo {author} {\bibfnamefont {D.}~\bibnamefont {Makarov}},
  \bibinfo {author} {\bibfnamefont {M.}~\bibnamefont {Albrecht}}, \bibinfo
  {author} {\bibfnamefont {N.}~\bibnamefont {Rivier}}, \bibinfo {author}
  {\bibfnamefont {P.}~\bibnamefont {Leiderer}},\ and\ \bibinfo {author}
  {\bibfnamefont {A.}~\bibnamefont {Erbe}},\ }\bibfield  {title} {\enquote
  {\bibinfo {title} {Frustration-induced magic number clusters of colloidal
  magnetic particles},}\ }\href {https://doi.org/10.1103/PhysRevE.77.031407}
  {\bibfield  {journal} {\bibinfo  {journal} {Physical Review E}\ }\textbf
  {\bibinfo {volume} {77}},\ \bibinfo {pages} {031407} (\bibinfo {year}
  {2008})}\BibitemShut {NoStop}%
\bibitem [{\citenamefont {Kantorovich}\ \emph
  {et~al.}(2011{\natexlab{a}})\citenamefont {Kantorovich}, \citenamefont
  {Weeber}, \citenamefont {Cerd{\`a}},\ and\ \citenamefont
  {Holm}}]{kantorovich2011}%
  \BibitemOpen
  \bibfield  {author} {\bibinfo {author} {\bibfnamefont {S.}~\bibnamefont
  {Kantorovich}}, \bibinfo {author} {\bibfnamefont {R.}~\bibnamefont {Weeber}},
  \bibinfo {author} {\bibfnamefont {J.~J.}\ \bibnamefont {Cerd{\`a}}},\ and\
  \bibinfo {author} {\bibfnamefont {C.}~\bibnamefont {Holm}},\ }\bibfield
  {title} {\enquote {\bibinfo {title} {Magnetic particles with shifted
  dipoles},}\ }\href {https://doi.org/10.1016/j.jmmm.2010.11.019} {\bibfield
  {journal} {\bibinfo  {journal} {Journal of Magnetism and Magnetic Materials}\
  }\bibinfo {series} {Proceedings of 12th {{International Conference}} on
  {{Magnetic Fluid}}},\ \textbf {\bibinfo {volume} {323}},\ \bibinfo {pages}
  {1269--1272} (\bibinfo {year} {2011}{\natexlab{a}})}\BibitemShut {NoStop}%
\bibitem [{\citenamefont {Tan}\ \emph {et~al.}(2024)\citenamefont {Tan},
  \citenamefont {Song}, \citenamefont {Wan}, \citenamefont {Huang},
  \citenamefont {Chai},\ and\ \citenamefont {Yang}}]{tan2024}%
  \BibitemOpen
  \bibfield  {author} {\bibinfo {author} {\bibfnamefont {X.}~\bibnamefont
  {Tan}}, \bibinfo {author} {\bibfnamefont {Y.}~\bibnamefont {Song}}, \bibinfo
  {author} {\bibfnamefont {C.}~\bibnamefont {Wan}}, \bibinfo {author}
  {\bibfnamefont {C.}~\bibnamefont {Huang}}, \bibinfo {author} {\bibfnamefont
  {Y.}~\bibnamefont {Chai}},\ and\ \bibinfo {author} {\bibfnamefont
  {Z.}~\bibnamefont {Yang}},\ }\bibfield  {title} {\enquote {\bibinfo {title}
  {Magnetic {{Janus Particles}}: {{Synthesis}} and {{Multifunctional
  Applications}}},}\ }\href {https://doi.org/10.1002/marc.202400866} {\bibfield
   {journal} {\bibinfo  {journal} {Macromolecular Rapid Communications}\
  }\textbf {\bibinfo {volume} {n/a}},\ \bibinfo {pages} {2400866} (\bibinfo
  {year} {2024})}\BibitemShut {NoStop}%
\bibitem [{\citenamefont {Han}\ \emph {et~al.}(2017)\citenamefont {Han},
  \citenamefont {Shields}, \citenamefont {Diwakar}, \citenamefont {Bharti},
  \citenamefont {L{\'o}pez},\ and\ \citenamefont {Velev}}]{han2017b}%
  \BibitemOpen
  \bibfield  {author} {\bibinfo {author} {\bibfnamefont {K.}~\bibnamefont
  {Han}}, \bibinfo {author} {\bibfnamefont {C.~W.}\ \bibnamefont {Shields}},
  \bibinfo {author} {\bibfnamefont {N.~M.}\ \bibnamefont {Diwakar}}, \bibinfo
  {author} {\bibfnamefont {B.}~\bibnamefont {Bharti}}, \bibinfo {author}
  {\bibfnamefont {G.~P.}\ \bibnamefont {L{\'o}pez}},\ and\ \bibinfo {author}
  {\bibfnamefont {O.~D.}\ \bibnamefont {Velev}},\ }\bibfield  {title} {\enquote
  {\bibinfo {title} {Sequence-encoded colloidal origami and microbot assemblies
  from patchy magnetic cubes},}\ }\href
  {https://doi.org/10.1126/sciadv.1701108} {\bibfield  {journal} {\bibinfo
  {journal} {Science Advances}\ }\textbf {\bibinfo {volume} {3}},\ \bibinfo
  {pages} {e1701108} (\bibinfo {year} {2017})}\BibitemShut {NoStop}%
\bibitem [{\citenamefont {Long}, \citenamefont {{C{\'o}rdova-Figueroa}},\ and\
  \citenamefont {Kretzschmar}(2019)}]{long2019}%
  \BibitemOpen
  \bibfield  {author} {\bibinfo {author} {\bibfnamefont {T.~W.}\ \bibnamefont
  {Long}}, \bibinfo {author} {\bibfnamefont {U.~M.}\ \bibnamefont
  {{C{\'o}rdova-Figueroa}}},\ and\ \bibinfo {author} {\bibfnamefont
  {I.}~\bibnamefont {Kretzschmar}},\ }\bibfield  {title} {\enquote {\bibinfo
  {title} {Measuring, {{Modeling}}, and {{Predicting}} the {{Magnetic Assembly
  Rate}} of {{2D-Staggered Janus Particle Chains}}},}\ }\href
  {https://doi.org/10.1021/acs.langmuir.9b00163} {\bibfield  {journal}
  {\bibinfo  {journal} {Langmuir}\ }\textbf {\bibinfo {volume} {35}},\ \bibinfo
  {pages} {8121--8130} (\bibinfo {year} {2019})}\BibitemShut {NoStop}%
\bibitem [{\citenamefont {Philipp}\ \emph {et~al.}(2021)\citenamefont
  {Philipp}, \citenamefont {Gross}, \citenamefont {Reginka}, \citenamefont
  {Merkel}, \citenamefont {Claus}, \citenamefont {Sulliger}, \citenamefont
  {Ehresmann},\ and\ \citenamefont {Poggio}}]{philipp2021}%
  \BibitemOpen
  \bibfield  {author} {\bibinfo {author} {\bibfnamefont {S.}~\bibnamefont
  {Philipp}}, \bibinfo {author} {\bibfnamefont {B.}~\bibnamefont {Gross}},
  \bibinfo {author} {\bibfnamefont {M.}~\bibnamefont {Reginka}}, \bibinfo
  {author} {\bibfnamefont {M.}~\bibnamefont {Merkel}}, \bibinfo {author}
  {\bibfnamefont {M.~M.}\ \bibnamefont {Claus}}, \bibinfo {author}
  {\bibfnamefont {M.}~\bibnamefont {Sulliger}}, \bibinfo {author}
  {\bibfnamefont {A.}~\bibnamefont {Ehresmann}},\ and\ \bibinfo {author}
  {\bibfnamefont {M.}~\bibnamefont {Poggio}},\ }\bibfield  {title} {\enquote
  {\bibinfo {title} {Magnetic hysteresis of individual {{Janus}} particles with
  hemispherical exchange biased caps},}\ }\href
  {https://doi.org/10.1063/5.0076116} {\bibfield  {journal} {\bibinfo
  {journal} {Applied Physics Letters}\ }\textbf {\bibinfo {volume} {119}},\
  \bibinfo {pages} {222406} (\bibinfo {year} {2021})}\BibitemShut {NoStop}%
\bibitem [{\citenamefont {Eslami}, \citenamefont {Khanjari},\ and\
  \citenamefont {{M{\"u}ller-Plathe}}(2019)}]{eslami2019a}%
  \BibitemOpen
  \bibfield  {author} {\bibinfo {author} {\bibfnamefont {H.}~\bibnamefont
  {Eslami}}, \bibinfo {author} {\bibfnamefont {N.}~\bibnamefont {Khanjari}},\
  and\ \bibinfo {author} {\bibfnamefont {F.}~\bibnamefont
  {{M{\"u}ller-Plathe}}},\ }\bibfield  {title} {\enquote {\bibinfo {title}
  {Self-{{Assembly Mechanisms}} of {{Triblock Janus Particles}}},}\ }\href
  {https://doi.org/10.1021/acs.jctc.8b00713} {\bibfield  {journal} {\bibinfo
  {journal} {Journal of Chemical Theory and Computation}\ }\textbf {\bibinfo
  {volume} {15}},\ \bibinfo {pages} {1345--1354} (\bibinfo {year}
  {2019})}\BibitemShut {NoStop}%
\bibitem [{\citenamefont {Donaldson}, \citenamefont {Schall},\ and\
  \citenamefont {Rossi}(2021)}]{donaldson2021}%
  \BibitemOpen
  \bibfield  {author} {\bibinfo {author} {\bibfnamefont {J.~G.}\ \bibnamefont
  {Donaldson}}, \bibinfo {author} {\bibfnamefont {P.}~\bibnamefont {Schall}},\
  and\ \bibinfo {author} {\bibfnamefont {L.}~\bibnamefont {Rossi}},\ }\bibfield
   {title} {\enquote {\bibinfo {title} {Magnetic {{Coupling}} in {{Colloidal
  Clusters}} for {{Hierarchical Self-Assembly}}},}\ }\href
  {https://doi.org/10.1021/acsnano.0c09952} {\bibfield  {journal} {\bibinfo
  {journal} {ACS Nano}\ }\textbf {\bibinfo {volume} {15}},\ \bibinfo {pages}
  {4989--4999} (\bibinfo {year} {2021})}\BibitemShut {NoStop}%
\bibitem [{\citenamefont {Kantorovich}\ \emph
  {et~al.}(2011{\natexlab{b}})\citenamefont {Kantorovich}, \citenamefont
  {Weeber}, \citenamefont {Cerda},\ and\ \citenamefont
  {Holm}}]{kantorovich2011a}%
  \BibitemOpen
  \bibfield  {author} {\bibinfo {author} {\bibfnamefont {S.}~\bibnamefont
  {Kantorovich}}, \bibinfo {author} {\bibfnamefont {R.}~\bibnamefont {Weeber}},
  \bibinfo {author} {\bibfnamefont {J.~J.}\ \bibnamefont {Cerda}},\ and\
  \bibinfo {author} {\bibfnamefont {C.}~\bibnamefont {Holm}},\ }\bibfield
  {title} {\enquote {\bibinfo {title} {Ferrofluids with shifted dipoles: Ground
  state structures},}\ }\href {https://doi.org/10.1039/c1sm05186e} {\bibfield
  {journal} {\bibinfo  {journal} {Soft Matter}\ }\textbf {\bibinfo {volume}
  {7}},\ \bibinfo {pages} {5217} (\bibinfo {year}
  {2011}{\natexlab{b}})}\BibitemShut {NoStop}%
\bibitem [{\citenamefont {{Vega-Bellido}}\ \emph {et~al.}(2019)\citenamefont
  {{Vega-Bellido}}, \citenamefont {{DeLaCruz-Araujo}}, \citenamefont
  {Kretzschmar},\ and\ \citenamefont
  {{C{\'o}rdova-Figueroa}}}]{vega-bellido2019}%
  \BibitemOpen
  \bibfield  {author} {\bibinfo {author} {\bibfnamefont {G.~I.}\ \bibnamefont
  {{Vega-Bellido}}}, \bibinfo {author} {\bibfnamefont {R.~A.}\ \bibnamefont
  {{DeLaCruz-Araujo}}}, \bibinfo {author} {\bibfnamefont {I.}~\bibnamefont
  {Kretzschmar}},\ and\ \bibinfo {author} {\bibfnamefont {U.~M.}\ \bibnamefont
  {{C{\'o}rdova-Figueroa}}},\ }\bibfield  {title} {\enquote {\bibinfo {title}
  {Self-assembly of magnetic colloids with shifted dipoles},}\ }\href
  {https://doi.org/10.1039/C8SM02591F} {\bibfield  {journal} {\bibinfo
  {journal} {Soft Matter}\ }\textbf {\bibinfo {volume} {15}},\ \bibinfo {pages}
  {4078--4086} (\bibinfo {year} {2019})}\BibitemShut {NoStop}%
\bibitem [{\citenamefont {{A.~Victoria-Camacho}}\ \emph
  {et~al.}(2020)\citenamefont {{A.~Victoria-Camacho}}, \citenamefont
  {{A.~DeLaCruz-Araujo}}, \citenamefont {Kretzschmar},\ and\ \citenamefont
  {{M.~C{\'o}rdova-Figueroa}}}]{a.victoria-camacho2020}%
  \BibitemOpen
  \bibfield  {author} {\bibinfo {author} {\bibfnamefont {J.}~\bibnamefont
  {{A.~Victoria-Camacho}}}, \bibinfo {author} {\bibfnamefont {R.}~\bibnamefont
  {{A.~DeLaCruz-Araujo}}}, \bibinfo {author} {\bibfnamefont {I.}~\bibnamefont
  {Kretzschmar}},\ and\ \bibinfo {author} {\bibfnamefont {U.}~\bibnamefont
  {{M.~C{\'o}rdova-Figueroa}}},\ }\bibfield  {title} {\enquote {\bibinfo
  {title} {Self-assembly of magnetic colloids with radially shifted dipoles},}\
  }\href {https://doi.org/10.1039/C9SM02020A} {\bibfield  {journal} {\bibinfo
  {journal} {Soft Matter}\ }\textbf {\bibinfo {volume} {16}},\ \bibinfo {pages}
  {2460--2472} (\bibinfo {year} {2020})}\BibitemShut {NoStop}%
\bibitem [{\citenamefont {{A.~DeLaCruz-Araujo}}\ \emph
  {et~al.}(2016)\citenamefont {{A.~DeLaCruz-Araujo}}, \citenamefont
  {{J.~Beltran-Villegas}}, \citenamefont {G.~Larson},\ and\ \citenamefont
  {{M.~C{\'o}rdova-Figueroa}}}]{a.delacruz-araujo2016a}%
  \BibitemOpen
  \bibfield  {author} {\bibinfo {author} {\bibfnamefont {R.}~\bibnamefont
  {{A.~DeLaCruz-Araujo}}}, \bibinfo {author} {\bibfnamefont {D.}~\bibnamefont
  {{J.~Beltran-Villegas}}}, \bibinfo {author} {\bibfnamefont {R.}~\bibnamefont
  {G.~Larson}},\ and\ \bibinfo {author} {\bibfnamefont {U.}~\bibnamefont
  {{M.~C{\'o}rdova-Figueroa}}},\ }\bibfield  {title} {\enquote {\bibinfo
  {title} {Rich {{Janus}} colloid phase behavior under steady shear},}\ }\href
  {https://doi.org/10.1039/C6SM00183A} {\bibfield  {journal} {\bibinfo
  {journal} {Soft Matter}\ }\textbf {\bibinfo {volume} {12}},\ \bibinfo {pages}
  {4071--4081} (\bibinfo {year} {2016})}\BibitemShut {NoStop}%
\bibitem [{\citenamefont {Deaven}\ and\ \citenamefont {Ho}(1995)}]{deaven1995}%
  \BibitemOpen
  \bibfield  {author} {\bibinfo {author} {\bibfnamefont {D.~M.}\ \bibnamefont
  {Deaven}}\ and\ \bibinfo {author} {\bibfnamefont {K.~M.}\ \bibnamefont
  {Ho}},\ }\bibfield  {title} {\enquote {\bibinfo {title} {Molecular {{Geometry
  Optimization}} with a {{Genetic Algorithm}}},}\ }\href
  {https://doi.org/10.1103/PhysRevLett.75.288} {\bibfield  {journal} {\bibinfo
  {journal} {Physical Review Letters}\ }\textbf {\bibinfo {volume} {75}},\
  \bibinfo {pages} {288--291} (\bibinfo {year} {1995})}\BibitemShut {NoStop}%
\bibitem [{\citenamefont {Daven}\ \emph {et~al.}(1996)\citenamefont {Daven},
  \citenamefont {Tit}, \citenamefont {Morris},\ and\ \citenamefont
  {Ho}}]{daven1996}%
  \BibitemOpen
  \bibfield  {author} {\bibinfo {author} {\bibfnamefont {D.~M.}\ \bibnamefont
  {Daven}}, \bibinfo {author} {\bibfnamefont {N.}~\bibnamefont {Tit}}, \bibinfo
  {author} {\bibfnamefont {J.~R.}\ \bibnamefont {Morris}},\ and\ \bibinfo
  {author} {\bibfnamefont {K.~M.}\ \bibnamefont {Ho}},\ }\bibfield  {title}
  {\enquote {\bibinfo {title} {Structural optimization of {{Lennard-Jones}}
  clusters by a genetic algorithm},}\ }\href
  {https://doi.org/10.1016/0009-2614(96)00406-X} {\bibfield  {journal}
  {\bibinfo  {journal} {Chemical Physics Letters}\ }\textbf {\bibinfo {volume}
  {256}},\ \bibinfo {pages} {195--200} (\bibinfo {year} {1996})}\BibitemShut
  {NoStop}%
\bibitem [{\citenamefont {Chremos}\ and\ \citenamefont
  {Likos}(2009)}]{chremos2009}%
  \BibitemOpen
  \bibfield  {author} {\bibinfo {author} {\bibfnamefont {A.}~\bibnamefont
  {Chremos}}\ and\ \bibinfo {author} {\bibfnamefont {C.~N.}\ \bibnamefont
  {Likos}},\ }\bibfield  {title} {\enquote {\bibinfo {title} {Crystal
  {{Structures}} of {{Two-Dimensional Binary Mixtures}} of {{Dipolar Colloids}}
  in {{Tilted External Magnetic Fields}}},}\ }\href
  {https://doi.org/10.1021/jp903820d} {\bibfield  {journal} {\bibinfo
  {journal} {The Journal of Physical Chemistry B}\ }\textbf {\bibinfo {volume}
  {113}},\ \bibinfo {pages} {12316--12325} (\bibinfo {year}
  {2009})}\BibitemShut {NoStop}%
\bibitem [{\citenamefont {Doppelbauer}, \citenamefont {Bianchi},\ and\
  \citenamefont {Kahl}(2010)}]{doppelbauer2010}%
  \BibitemOpen
  \bibfield  {author} {\bibinfo {author} {\bibfnamefont {G.}~\bibnamefont
  {Doppelbauer}}, \bibinfo {author} {\bibfnamefont {E.}~\bibnamefont
  {Bianchi}},\ and\ \bibinfo {author} {\bibfnamefont {G.}~\bibnamefont
  {Kahl}},\ }\bibfield  {title} {\enquote {\bibinfo {title} {Self-assembly
  scenarios of patchy colloidal particles in two dimensions},}\ }\href
  {https://doi.org/10.1088/0953-8984/22/10/104105} {\bibfield  {journal}
  {\bibinfo  {journal} {Journal of Physics: Condensed Matter}\ }\textbf
  {\bibinfo {volume} {22}},\ \bibinfo {pages} {104105} (\bibinfo {year}
  {2010})}\BibitemShut {NoStop}%
\bibitem [{\citenamefont {Bianchi}\ \emph {et~al.}(2012)\citenamefont
  {Bianchi}, \citenamefont {Doppelbauer}, \citenamefont {Filion}, \citenamefont
  {Dijkstra},\ and\ \citenamefont {Kahl}}]{bianchi2012}%
  \BibitemOpen
  \bibfield  {author} {\bibinfo {author} {\bibfnamefont {E.}~\bibnamefont
  {Bianchi}}, \bibinfo {author} {\bibfnamefont {G.}~\bibnamefont
  {Doppelbauer}}, \bibinfo {author} {\bibfnamefont {L.}~\bibnamefont {Filion}},
  \bibinfo {author} {\bibfnamefont {M.}~\bibnamefont {Dijkstra}},\ and\
  \bibinfo {author} {\bibfnamefont {G.}~\bibnamefont {Kahl}},\ }\bibfield
  {title} {\enquote {\bibinfo {title} {Predicting patchy particle crystals:
  {{Variable}} box shape simulations and evolutionary algorithms},}\ }\href
  {https://doi.org/10.1063/1.4722477} {\bibfield  {journal} {\bibinfo
  {journal} {The Journal of Chemical Physics}\ }\textbf {\bibinfo {volume}
  {136}},\ \bibinfo {pages} {214102} (\bibinfo {year} {2012})}\BibitemShut
  {NoStop}%
\bibitem [{\citenamefont {Cruz}, \citenamefont {Marques},\ and\ \citenamefont
  {Pereira}(2016)}]{cruz2016a}%
  \BibitemOpen
  \bibfield  {author} {\bibinfo {author} {\bibfnamefont {S.~M.~A.}\
  \bibnamefont {Cruz}}, \bibinfo {author} {\bibfnamefont {J.~M.~C.}\
  \bibnamefont {Marques}},\ and\ \bibinfo {author} {\bibfnamefont {F.~B.}\
  \bibnamefont {Pereira}},\ }\bibfield  {title} {\enquote {\bibinfo {title}
  {Improved evolutionary algorithm for the global optimization of clusters with
  competing attractive and repulsive interactions},}\ }\href
  {https://doi.org/10.1063/1.4964780} {\bibfield  {journal} {\bibinfo
  {journal} {The Journal of Chemical Physics}\ }\textbf {\bibinfo {volume}
  {145}},\ \bibinfo {pages} {154109} (\bibinfo {year} {2016})}\BibitemShut
  {NoStop}%
\bibitem [{\citenamefont {Sch{\"o}nborn}\ \emph {et~al.}(2009)\citenamefont
  {Sch{\"o}nborn}, \citenamefont {Goedecker}, \citenamefont {Roy},\ and\
  \citenamefont {Oganov}}]{schonborn2009}%
  \BibitemOpen
  \bibfield  {author} {\bibinfo {author} {\bibfnamefont {S.~E.}\ \bibnamefont
  {Sch{\"o}nborn}}, \bibinfo {author} {\bibfnamefont {S.}~\bibnamefont
  {Goedecker}}, \bibinfo {author} {\bibfnamefont {S.}~\bibnamefont {Roy}},\
  and\ \bibinfo {author} {\bibfnamefont {A.~R.}\ \bibnamefont {Oganov}},\
  }\bibfield  {title} {\enquote {\bibinfo {title} {The performance of minima
  hopping and evolutionary algorithms for cluster structure prediction},}\
  }\href {https://doi.org/10.1063/1.3097197} {\bibfield  {journal} {\bibinfo
  {journal} {The Journal of Chemical Physics}\ }\textbf {\bibinfo {volume}
  {130}},\ \bibinfo {pages} {144108} (\bibinfo {year} {2009})}\BibitemShut
  {NoStop}%
\bibitem [{\citenamefont {Brown}, \citenamefont {Thompson},\ and\ \citenamefont
  {Schultz}(2010)}]{brown2010}%
  \BibitemOpen
  \bibfield  {author} {\bibinfo {author} {\bibfnamefont {W.~M.}\ \bibnamefont
  {Brown}}, \bibinfo {author} {\bibfnamefont {A.~P.}\ \bibnamefont
  {Thompson}},\ and\ \bibinfo {author} {\bibfnamefont {P.~A.}\ \bibnamefont
  {Schultz}},\ }\bibfield  {title} {\enquote {\bibinfo {title} {Efficient
  hybrid evolutionary optimization of interatomic potential models},}\ }\href
  {https://doi.org/10.1063/1.3294562} {\bibfield  {journal} {\bibinfo
  {journal} {The Journal of Chemical Physics}\ }\textbf {\bibinfo {volume}
  {132}},\ \bibinfo {pages} {024108} (\bibinfo {year} {2010})}\BibitemShut
  {NoStop}%
\bibitem [{\citenamefont {Storn}\ and\ \citenamefont
  {Price}(1997)}]{storn1997}%
  \BibitemOpen
  \bibfield  {author} {\bibinfo {author} {\bibfnamefont {R.}~\bibnamefont
  {Storn}}\ and\ \bibinfo {author} {\bibfnamefont {K.}~\bibnamefont {Price}},\
  }\bibfield  {title} {\enquote {\bibinfo {title} {Differential {{Evolution}}
  -- {{A Simple}} and {{Efficient Heuristic}} for global {{Optimization}} over
  {{Continuous Spaces}}},}\ }\href {https://doi.org/10.1023/A:1008202821328}
  {\bibfield  {journal} {\bibinfo  {journal} {Journal of Global Optimization}\
  }\textbf {\bibinfo {volume} {11}},\ \bibinfo {pages} {341--359} (\bibinfo
  {year} {1997})}\BibitemShut {NoStop}%
\bibitem [{\citenamefont {Ahmad}\ \emph {et~al.}(2022)\citenamefont {Ahmad},
  \citenamefont {Isa}, \citenamefont {Lim},\ and\ \citenamefont
  {Ang}}]{ahmad2022}%
  \BibitemOpen
  \bibfield  {author} {\bibinfo {author} {\bibfnamefont {M.~F.}\ \bibnamefont
  {Ahmad}}, \bibinfo {author} {\bibfnamefont {N.~A.~M.}\ \bibnamefont {Isa}},
  \bibinfo {author} {\bibfnamefont {W.~H.}\ \bibnamefont {Lim}},\ and\ \bibinfo
  {author} {\bibfnamefont {K.~M.}\ \bibnamefont {Ang}},\ }\bibfield  {title}
  {\enquote {\bibinfo {title} {Differential evolution: {{A}} recent review
  based on state-of-the-art works},}\ }\href
  {https://doi.org/10.1016/j.aej.2021.09.013} {\bibfield  {journal} {\bibinfo
  {journal} {Alexandria Engineering Journal}\ }\textbf {\bibinfo {volume}
  {61}},\ \bibinfo {pages} {3831--3872} (\bibinfo {year} {2022})}\BibitemShut
  {NoStop}%
\bibitem [{\citenamefont {Moloi}\ and\ \citenamefont {Ali}(2005)}]{moloi2005}%
  \BibitemOpen
  \bibfield  {author} {\bibinfo {author} {\bibfnamefont {N.~P.}\ \bibnamefont
  {Moloi}}\ and\ \bibinfo {author} {\bibfnamefont {M.~M.}\ \bibnamefont
  {Ali}},\ }\bibfield  {title} {\enquote {\bibinfo {title} {An {{Iterative
  Global Optimization Algorithm}} for {{Potential Energy Minimization}}},}\
  }\href {https://doi.org/10.1007/s10589-005-4555-9} {\bibfield  {journal}
  {\bibinfo  {journal} {Computational Optimization and Applications}\ }\textbf
  {\bibinfo {volume} {30}},\ \bibinfo {pages} {119--132} (\bibinfo {year}
  {2005})}\BibitemShut {NoStop}%
\bibitem [{\citenamefont {Okada}\ and\ \citenamefont
  {Satoh}(2023)}]{okada2023}%
  \BibitemOpen
  \bibfield  {author} {\bibinfo {author} {\bibfnamefont {K.}~\bibnamefont
  {Okada}}\ and\ \bibinfo {author} {\bibfnamefont {A.}~\bibnamefont {Satoh}},\
  }\bibfield  {title} {\enquote {\bibinfo {title} {Regime change in the
  aggregate structures of spherical magnetic {{Janus}} particles (quasi-{{2D
  Monte Carlo}} simulations)},}\ }\href
  {https://doi.org/10.1080/00268976.2023.2245503} {\bibfield  {journal}
  {\bibinfo  {journal} {Molecular Physics}\ }\textbf {\bibinfo {volume}
  {121}},\ \bibinfo {pages} {e2245503} (\bibinfo {year} {2023})}\BibitemShut
  {NoStop}%
\bibitem [{\citenamefont {Novak}, \citenamefont {Pyanzina},\ and\ \citenamefont
  {Kantorovich}(2015)}]{novak2015b}%
  \BibitemOpen
  \bibfield  {author} {\bibinfo {author} {\bibfnamefont {E.~V.}\ \bibnamefont
  {Novak}}, \bibinfo {author} {\bibfnamefont {E.~S.}\ \bibnamefont
  {Pyanzina}},\ and\ \bibinfo {author} {\bibfnamefont {S.~S.}\ \bibnamefont
  {Kantorovich}},\ }\bibfield  {title} {\enquote {\bibinfo {title} {Behaviour
  of magnetic {{Janus-like}} colloids},}\ }\href
  {https://doi.org/10.1088/0953-8984/27/23/234102} {\bibfield  {journal}
  {\bibinfo  {journal} {Journal of Physics: Condensed Matter}\ }\textbf
  {\bibinfo {volume} {27}},\ \bibinfo {pages} {234102} (\bibinfo {year}
  {2015})}\BibitemShut {NoStop}%
\bibitem [{\citenamefont {Cristina}(2021)}]{cristina2021}%
  \BibitemOpen
  \bibfield  {author} {\bibinfo {author} {\bibfnamefont {S.}~\bibnamefont
  {Cristina}},\ }\href@noop {} {\enquote {\bibinfo {title} {Differential
  {{Evolution}} from {{Scratch}} in {{Python}}},}\ }\bibinfo {howpublished}
  {https://www.machinelearningmastery.com/differential-evolution-from-scratch-in-python/}
  (\bibinfo {year} {2021})\BibitemShut {NoStop}%
\bibitem [{\citenamefont {Harris}\ \emph {et~al.}(2020)\citenamefont {Harris},
  \citenamefont {Millman}, \citenamefont {Van Der~Walt}, \citenamefont
  {Gommers}, \citenamefont {Virtanen}, \citenamefont {Cournapeau},
  \citenamefont {Wieser}, \citenamefont {Taylor}, \citenamefont {Berg},
  \citenamefont {Smith}, \citenamefont {Kern}, \citenamefont {Picus},
  \citenamefont {Hoyer}, \citenamefont {Van~Kerkwijk}, \citenamefont {Brett},
  \citenamefont {Haldane}, \citenamefont {Del~R{\'i}o}, \citenamefont {Wiebe},
  \citenamefont {Peterson}, \citenamefont {{G{\'e}rard-Marchant}},
  \citenamefont {Sheppard}, \citenamefont {Reddy}, \citenamefont {Weckesser},
  \citenamefont {Abbasi}, \citenamefont {Gohlke},\ and\ \citenamefont
  {Oliphant}}]{harris2020}%
  \BibitemOpen
  \bibfield  {author} {\bibinfo {author} {\bibfnamefont {C.~R.}\ \bibnamefont
  {Harris}}, \bibinfo {author} {\bibfnamefont {K.~J.}\ \bibnamefont {Millman}},
  \bibinfo {author} {\bibfnamefont {S.~J.}\ \bibnamefont {Van Der~Walt}},
  \bibinfo {author} {\bibfnamefont {R.}~\bibnamefont {Gommers}}, \bibinfo
  {author} {\bibfnamefont {P.}~\bibnamefont {Virtanen}}, \bibinfo {author}
  {\bibfnamefont {D.}~\bibnamefont {Cournapeau}}, \bibinfo {author}
  {\bibfnamefont {E.}~\bibnamefont {Wieser}}, \bibinfo {author} {\bibfnamefont
  {J.}~\bibnamefont {Taylor}}, \bibinfo {author} {\bibfnamefont
  {S.}~\bibnamefont {Berg}}, \bibinfo {author} {\bibfnamefont {N.~J.}\
  \bibnamefont {Smith}}, \bibinfo {author} {\bibfnamefont {R.}~\bibnamefont
  {Kern}}, \bibinfo {author} {\bibfnamefont {M.}~\bibnamefont {Picus}},
  \bibinfo {author} {\bibfnamefont {S.}~\bibnamefont {Hoyer}}, \bibinfo
  {author} {\bibfnamefont {M.~H.}\ \bibnamefont {Van~Kerkwijk}}, \bibinfo
  {author} {\bibfnamefont {M.}~\bibnamefont {Brett}}, \bibinfo {author}
  {\bibfnamefont {A.}~\bibnamefont {Haldane}}, \bibinfo {author} {\bibfnamefont
  {J.~F.}\ \bibnamefont {Del~R{\'i}o}}, \bibinfo {author} {\bibfnamefont
  {M.}~\bibnamefont {Wiebe}}, \bibinfo {author} {\bibfnamefont
  {P.}~\bibnamefont {Peterson}}, \bibinfo {author} {\bibfnamefont
  {P.}~\bibnamefont {{G{\'e}rard-Marchant}}}, \bibinfo {author} {\bibfnamefont
  {K.}~\bibnamefont {Sheppard}}, \bibinfo {author} {\bibfnamefont
  {T.}~\bibnamefont {Reddy}}, \bibinfo {author} {\bibfnamefont
  {W.}~\bibnamefont {Weckesser}}, \bibinfo {author} {\bibfnamefont
  {H.}~\bibnamefont {Abbasi}}, \bibinfo {author} {\bibfnamefont
  {C.}~\bibnamefont {Gohlke}},\ and\ \bibinfo {author} {\bibfnamefont {T.~E.}\
  \bibnamefont {Oliphant}},\ }\bibfield  {title} {\enquote {\bibinfo {title}
  {Array programming with {{NumPy}}},}\ }\href
  {https://doi.org/10.1038/s41586-020-2649-2} {\bibfield  {journal} {\bibinfo
  {journal} {Nature}\ }\textbf {\bibinfo {volume} {585}},\ \bibinfo {pages}
  {357--362} (\bibinfo {year} {2020})}\BibitemShut {NoStop}%
\bibitem [{\citenamefont {Katoch}, \citenamefont {Chauhan},\ and\ \citenamefont
  {Kumar}(2021)}]{katoch2021b}%
  \BibitemOpen
  \bibfield  {author} {\bibinfo {author} {\bibfnamefont {S.}~\bibnamefont
  {Katoch}}, \bibinfo {author} {\bibfnamefont {S.~S.}\ \bibnamefont
  {Chauhan}},\ and\ \bibinfo {author} {\bibfnamefont {V.}~\bibnamefont
  {Kumar}},\ }\bibfield  {title} {\enquote {\bibinfo {title} {A review on
  genetic algorithm: Past, present, and future},}\ }\href
  {https://doi.org/10.1007/s11042-020-10139-6} {\bibfield  {journal} {\bibinfo
  {journal} {Multimedia Tools and Applications}\ }\textbf {\bibinfo {volume}
  {80}},\ \bibinfo {pages} {8091--8126} (\bibinfo {year} {2021})}\BibitemShut
  {NoStop}%
\bibitem [{\citenamefont {Zhang}(2023)}]{zhang2023}%
  \BibitemOpen
  \bibfield  {author} {\bibinfo {author} {\bibfnamefont {X.}~\bibnamefont
  {Zhang}},\ }\bibfield  {title} {\enquote {\bibinfo {title} {Differential
  {{Evolution}} without the {{Scale Factor}} and the {{Crossover
  Probability}}},}\ }\href {https://doi.org/10.1155/2023/8973912} {\bibfield
  {journal} {\bibinfo  {journal} {Journal of Mathematics}\ }\textbf {\bibinfo
  {volume} {2023}},\ \bibinfo {pages} {8973912} (\bibinfo {year}
  {2023})}\BibitemShut {NoStop}%
\bibitem [{\citenamefont {Botev}, \citenamefont {Grotowski},\ and\
  \citenamefont {Kroese}(2010)}]{botev2010}%
  \BibitemOpen
  \bibfield  {author} {\bibinfo {author} {\bibfnamefont {Z.~I.}\ \bibnamefont
  {Botev}}, \bibinfo {author} {\bibfnamefont {J.~F.}\ \bibnamefont
  {Grotowski}},\ and\ \bibinfo {author} {\bibfnamefont {D.~P.}\ \bibnamefont
  {Kroese}},\ }\bibfield  {title} {\enquote {\bibinfo {title} {Kernel density
  estimation via diffusion},}\ }\href {https://doi.org/10.1214/10-AOS799}
  {\bibfield  {journal} {\bibinfo  {journal} {The Annals of Statistics}\
  }\textbf {\bibinfo {volume} {38}},\ \bibinfo {pages} {2916--2957} (\bibinfo
  {year} {2010})}\BibitemShut {NoStop}%
\bibitem [{202(2025)}]{2025}%
  \BibitemOpen
  \href@noop {} {\enquote {\bibinfo {title} {{{KDEpy}} 1.1.11 documentation},}\
  }\bibinfo {howpublished} {https://kdepy.readthedocs.io/en/latest/} (\bibinfo
  {year} {2025})\BibitemShut {NoStop}%
\bibitem [{\citenamefont {Klinkigt}\ \emph {et~al.}(2013)\citenamefont
  {Klinkigt}, \citenamefont {Weeber}, \citenamefont {Kantorovich},\ and\
  \citenamefont {Holm}}]{klinkigt2013}%
  \BibitemOpen
  \bibfield  {author} {\bibinfo {author} {\bibfnamefont {M.}~\bibnamefont
  {Klinkigt}}, \bibinfo {author} {\bibfnamefont {R.}~\bibnamefont {Weeber}},
  \bibinfo {author} {\bibfnamefont {S.}~\bibnamefont {Kantorovich}},\ and\
  \bibinfo {author} {\bibfnamefont {C.}~\bibnamefont {Holm}},\ }\bibfield
  {title} {\enquote {\bibinfo {title} {Cluster formation in systems of
  shifted-dipole particles},}\ }\href {https://doi.org/10.1039/c2sm27290c}
  {\bibfield  {journal} {\bibinfo  {journal} {Soft Matter}\ }\textbf {\bibinfo
  {volume} {9}},\ \bibinfo {pages} {3535} (\bibinfo {year} {2013})}\BibitemShut
  {NoStop}%
\bibitem [{\citenamefont {Weeber}\ \emph {et~al.}(2013)\citenamefont {Weeber},
  \citenamefont {Klinkigt}, \citenamefont {Kantorovich},\ and\ \citenamefont
  {Holm}}]{weeber2013}%
  \BibitemOpen
  \bibfield  {author} {\bibinfo {author} {\bibfnamefont {R.}~\bibnamefont
  {Weeber}}, \bibinfo {author} {\bibfnamefont {M.}~\bibnamefont {Klinkigt}},
  \bibinfo {author} {\bibfnamefont {S.}~\bibnamefont {Kantorovich}},\ and\
  \bibinfo {author} {\bibfnamefont {C.}~\bibnamefont {Holm}},\ }\bibfield
  {title} {\enquote {\bibinfo {title} {Microstructure and magnetic properties
  of magnetic fluids consisting of shifted dipole particles under the influence
  of an external magnetic field},}\ }\href {https://doi.org/10.1063/1.4832239}
  {\bibfield  {journal} {\bibinfo  {journal} {The Journal of Chemical Physics}\
  }\textbf {\bibinfo {volume} {139}},\ \bibinfo {pages} {214901} (\bibinfo
  {year} {2013})}\BibitemShut {NoStop}%
\bibitem [{202(2024)}]{2024}%
  \BibitemOpen
  \href@noop {} {\enquote {\bibinfo {title} {{{PlotDigitizer}}: {{Version}}
  3.1.6},}\ }\bibinfo {howpublished} {https://plotdigitizer.com} (\bibinfo
  {year} {2024})\BibitemShut {NoStop}%
\bibitem [{\citenamefont {Biswas}(2017)}]{biswas2017}%
  \BibitemOpen
  \bibfield  {author} {\bibinfo {author} {\bibfnamefont {K.}~\bibnamefont
  {Biswas}},\ }\bibfield  {title} {\enquote {\bibinfo {title} {A thermally
  driven differential mutation approach for the structural optimization of
  large atomic systems},}\ }\href {https://doi.org/10.1063/1.4986303}
  {\bibfield  {journal} {\bibinfo  {journal} {The Journal of Chemical Physics}\
  }\textbf {\bibinfo {volume} {147}},\ \bibinfo {pages} {104108} (\bibinfo
  {year} {2017})}\BibitemShut {NoStop}%
\end{thebibliography}%

\end{document}